REVIEW

# A comprehensive survey of orbital edge computing: Systems, applications, and algorithms

Zengshan YIN [a,b], Changhao WU [a,b], Chongbin GUO [a,b,*], Yuanchun LI [c], Mengwei XU [d], Weiwei GAO [d], Chuanxiu CHI [d]

[a] Chinese Academy of Sciences Shanghai Innovation Academy for Microsatellites, Shanghai 201304, China
[b] University of Chinese Academy of Sciences, Beijing 101408, China
[c] Institute for AI Industry Research (AIR), Tsinghua University, Beijing 102200, China
[d] State Key Laboratory of Networking and Switching Technology Computer Science Department, Beijing University of Posts and Telecommunications, Beijing 100876, China

**Abstract** The number of satellites, especially those operating in Low-Earth Orbit (LEO), has been exploding in recent years. Additionally, the burgeoning development of Artificial Intelligence (AI) software and hardware has opened up new industrial opportunities in both air and space, with satellite-powered computing emerging as a new computing paradigm: Orbital Edge Computing (OEC). Compared to terrestrial edge computing, the mobility of LEO satellites and their limited communication, computation, and storage resources pose challenges in designing task-specific scheduling algorithms. Previous survey papers have largely focused on terrestrial edge computing or the integration of space and ground technologies, lacking a comprehensive summary of OEC architecture, algorithms, and case studies. This paper conducts a comprehensive survey and analysis of OEC's system architecture, applications, algorithms, and simulation tools, providing a solid background for researchers in the field. By discussing OEC use cases and the challenges faced, potential research directions for future OEC research are proposed.

## 1. Introduction

Low-Earth Orbit (LEO) satellite networks show promise due to their ubiquity and invulnerability.[1] Excellent performance in latency and coverage makes LEO satellites a promising platform for extending mobile communication networks and providing ubiquitous wireless access. Large-scale LEO satellite networks have been widely researched for their potential.[2] Currently, several LEO satellite projects aimed at providing global communications have been launched, such as OneWeb,[3] SpaceX Starlink,[4] Kuiper,[5] and Telesat.[6] These projects plan to build mega-constellations consisting of thousands of LEO satellites to enable computation-intensive services on the ground and in space.[1]

* Corresponding author at: Chinese Academy of Sciences Shanghai Innovation Academy for Microsatellites, Shanghai 201304, China.
E-mail address: guocb@microsate.com (C. GUO).

The development of Artificial Intelligence (AI) has changed various industries, leading to an unprecedented demand for computing. The massive computing load and high-performance communication requirements have driven the emergence of concepts such as cloud computing[7] and edge computing.[8] Computing has been bound with communication networks. In the last decade, the rapid development of satellite networks has made it possible to envision a computing network in space.[9]

Traditionally, spacecraft prioritized high reliability and safety, resulting in lower computing capabilities for satellites and limiting their applications to specific missions.[10] Today, the rapid development of LEO satellites and inter-satellite communication technologies has paved the way for the construction of space-based computing clusters. The need for large-scale satellite constellations necessitates batch production of satellites. This approach can significantly reduce manufacturing and launch costs, making computing services within the network a viable possibility.[11]

Based on the above feature of LEO satellites, OEC is a promising technology that can provide communication, navigation, and computation for marine users,[12] disaster area,[13] low-altitude economy,[14] and other remote area that terrestrial communications are hard to reach. In addition, remote sensing missions such as earth observation and deep space exploration[15] will generate a large amount of data every moment, which is limited by communication resources between satellites and Ground Stations (GSs) and requires space-based high-timeliness computing capabilities.[9]

Moreover, an LEO satellite network can also support computation services for users in remote areas and the ocean.[12] It is tough to deploy an MEC in a satellite network and meet the needs of emerging applications, such as Augmented Reality (AR)/Virtual Reality (VR), High Definition (HD) video transmission, and autonomous driving.[16] Furthermore, orbital tasks such as remote sensing and space situational awareness[15] generate a massive amount of data every moment, which demands powerful computational power and the mitigation of any bottlenecks in the communication between the ground and space.[9]

Denby et al.[17,18] proposed a concept of OEC and showed that OEC improves the efficiency of remote sensing image processing by avoiding redundant data transmission between satellites and the ground. They constructed an OEC architecture comprising three formation-flying satellites, efficiently utilizing the computational resources of these satellites and enhancing the performance of remote sensing image processing.

Currently, OEC is considered a promising technology, for example, the remote sensing data captured by LEO satellites can be processed directly by the OEC, thus reducing the amount of data that needs to be transmitted back to the ground. Industry and academia have investigated a wide range of issues related to OEC, including satellite networking,[2] research platform,[9] testbed,[19] architecture,[16,20] resource management[21] and computation offloading strategy.[22,23] The National Aeronautics and Space Administration (NASA) has identified OEC as the next frontier of innovation and has announced its significant application value in space situational awareness, traffic data management, and satellite IoT.[24] European Space Agency (ESA) developed the EdgeSAT project, which was completed successfully in 2020. This project has identified satellite communication connectivity scenarios benefiting from edge networking capabilities.[25]

There are several published surveys about edge computing in satellite networks. However, systematic and comprehensive surveys of OEC are still lacking. Table 1[11,20,26–28] compares the present survey with the closest related work, showing that only a few surveys treat usable toolkits, testbeds, and research platforms.

In this survey, we provide a comprehensive review of recent advancements in orbital edge computing, with a particular focus on OEC systems, algorithms, and applications. Unlike previous surveys that primarily focus on architectures, our work highlights the simulators and testbeds. It is noticeable that we systematically discuss the availability and challenges of satellite hardware and software in OEC scenarios and summarize the architectural paradigms of OEC. Furthermore, we introduce a novel classification scheme for OEC applications. Both current challenges and future prospects are discussed in this survey.

Fig. 1 presents the outline of this paper. Section 2 introduces the OEC concept, background, and advanced space technology and compares it with the MEC in terrestrial networks. Section 3 introduces the characteristics of the satellite-specific software and hardware, as well as the cloud-edge architecture of the OEC. Various applications of OEC are summarized in Section 4. Testbeds and experimental methods are treated in Section 5. Section 6 summarizes the algorithms in OEC, focusing on convex optimization, game theory, and deep learning methods. Lastly, Section 7 presents the key challenges and future research directions.

## 2. Orbital edge computing

We will now introduce the concept and background of OEC. Then, we will introduce the satellite constellations and networking, which are the basis of the OEC architecture. Third, we introduce the evolution of OEC from ground-based mobile computing. Next, we compare OEC to terrestrial MEC. Finally, advanced technologies in space, such as laser communication, smart satellites, and the space-air-ground integrated network, are introduced.

### 2.1. Concept of OEC

OEC is an emerging paradigm that involves performing computational tasks directly on satellites in orbit instead of sending all the data back to Earth for processing. Satellites are equipped with powerful computing resources and connect and collaborate to construct OEC networks, which can provide computation offloading services for ground and airborne users.[29] Satellites and IoT devices (such as smartphones, tablets, monitors, etc.) can offload computing tasks to OEC networks to achieve high-speed, low-latency, and low-cost computing.

Traditionally, satellite networks are based on the bent-pipe architecture, which means satellites merely act as communication relay nodes to transmit data or as sensor nodes to receive instructions from the ground and then send the raw data back to the ground. In a massive constellation, this architecture can result in significant latency, complex routing, and other complications. Augmenting the computational capabilities of satellites has attracted considerable attention as a promising solution to this problem.[30] In the OEC architecture, satellites are equipped with computing capabilities, which can complete computing tasks as edge servers in the network.[31] Since the computational resources are located closer to the data sources, OEC greatly improves communication performance. This enables OEC to provide more diversified services, such as on-orbit data preprocessing, real-time remote sensing image analysis, and disaster monitoring.[32–34]

**Table 1** Comparison of present survey with closely related work.

| Reference | Satellite communication | Orbits and constellations for OEC | Hardware, software and system | Algorithms for OEC | Applications or use cases | Architecture | Toolkit | Testbed for OEC |
|---|---|---|---|---|---|---|---|---|
| 26 | | | | | ✓ | | ✓ | ✓ |
| 27 | | | | ✓ | ✓ | ✓ | | |
| 28 | ✓ | ✓ | | | ✓ | ✓ | | |
| 20 | | ✓ | | ✓ | | ✓ | | |
| 11 | ✓ | ✓ | | | ✓ | | | ✓ |
| Ours | ✓ | ✓ | ✓ | ✓ | ✓ | ✓ | ✓ | ✓ |

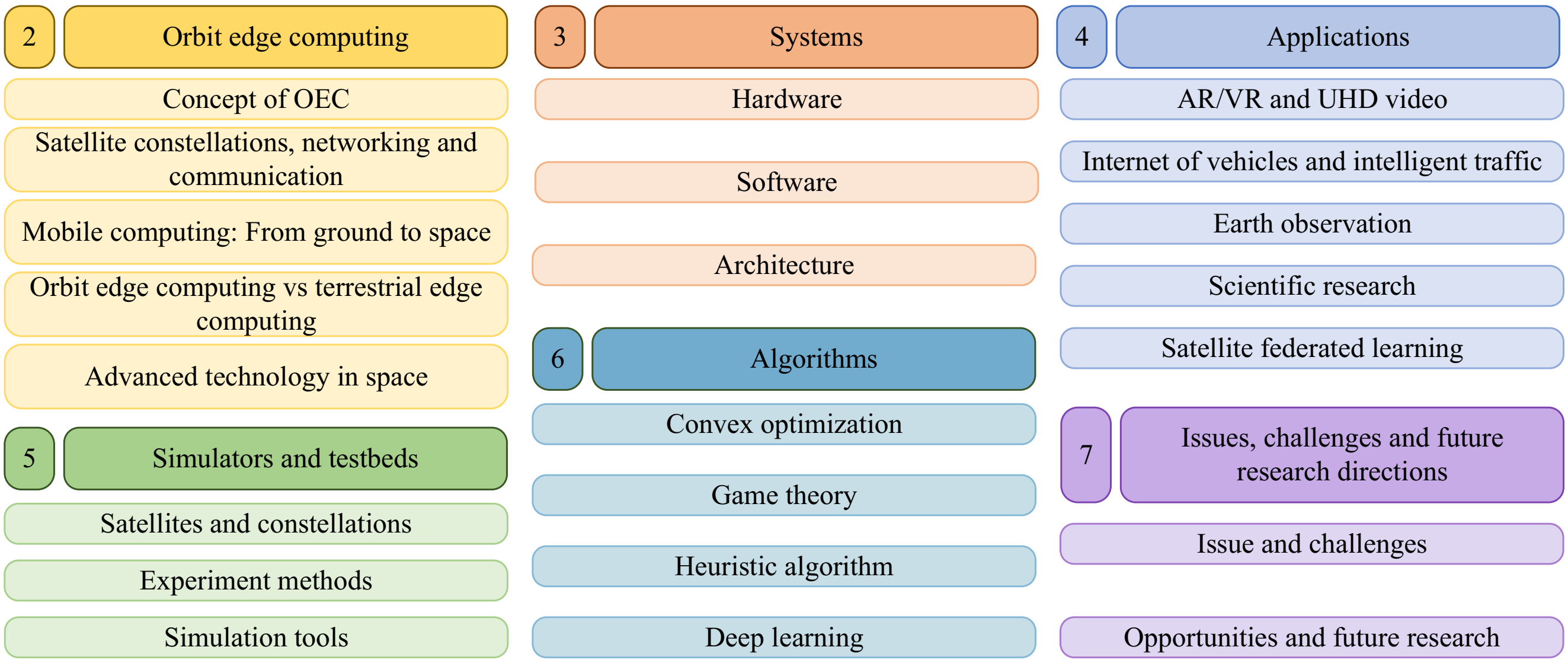

**Fig. 1** Outline of this paper.

### *2.2. Satellite constellations, networking and communication*

#### *2.2.1. Satellite constellation*

A satellite constellation is a system of artificial satellites with similar functions and parameters, such as orbital altitude and inclination. Hence, these satellites can be produced and launched in a batch, which leads to lower costs. They also offer unique communication and coverage performance that varies according to the orbital altitude. These distinctions are made clear in Table 2.[35] We introduce the differences between several types of satellite orbits.

Here, inclination refers to the angle between the orbital plane and the equatorial plane, while Altitude refers to the distance from the satellite in orbit to the Earth's surface. Note that elliptical orbits are described by their perigee (closest point to Earth) and apogee (farthest point from Earth). The altitudes of LEO, MEO, and GEO are described based on circular orbits, while the altitude range for HEO includes both perigee and apogee. Period refers to the time it takes for a satellite to orbit the Earth once, also known as the satellite's repeat period. This is different from the satellite's revisit period, which is the time it takes for the satellite to pass over the same position on the Earth's surface. This article does not use the revisit period as an orbital parameter. Number of satellites in a constellation refers to the general number of satellites deployed in a satellite constellation under the corresponding type of orbit. Coverage refers to the observable range of the constellation on Earth under the corresponding type of orbit. Latency refers to the transmission time of one-way communication from a satellite to a GS, which is generally related to the communication payload capacity of the satellite. Here, only typical values based on satellite orbital characteristics are listed. Pass durations denotes the time interval required for an individual satellite operating in a specific orbit to traverse the communication range of a GS.

**Table 2** Comparison of satellites in different orbits.[35]

| Orbit Type | Low Earth Orbit (LEO) | Highly Elliptical Orbit (HEO) | Middle Earth Orbit (MEO) | Geostationary Orbit (GEO) |
| --- | --- | --- | --- | --- |
| Inclination | 80 – 100° (near-polar orbit) 30 – 60° (inclined orbit) | 63.4°, 97.6° | 30 – 60° | 0° |
| Altitude | 300 – 1 500 km | 600 – 40 000 km | 8 000 – 20 000 km | 35 786 km |
| Period | 1.4 – 2.5 h | 12 h | 6 – 12 h | 24 h |
| Number of satellites in a constellation | 24 – thousands | 4 – 8 | 8 – 16 | 3 – 4 |
| Coverage | Global | High latitude areas | Global | Global (except polar regions) |
| Latency | 5 – 35 ms | 150 – 250 ms | 50 – 100 ms | 270 ms |
| Pass durations | About 10 min | 4 – 8 h | 1 – 2 h | All the time |
| Typical constellation | Iridium, Starlink, Kuiper, O3B | Molniya, Loopus, Archimedes | Odyssey | Inmarsat, MSAT Mobilesat |

[1] In summarizing the orbital inclination, we have only listed the orbital ranges of common communication, navigation, and remote sensing satellites. The ranges for specific requirements, such as meteorology or military uses, might not be on the list, and a special kind of orbit, the Sun-synchronous orbit, is also not on the list.

Sun-synchronous orbit is a special type of orbit with an inclination exceeding 90 degrees. Its advantage is the ability to provide stable sunlight conditions. Satellites in sun-synchronous orbits can observe clouds and ground targets under consistent time and lighting conditions.

Internet access based on GEO satellites has been available for decades: GEO satellites have the advantage of wide coverage, so global communication can be achieved with only three satellites (small regions around the North and South Poles are normally excluded). However, the high altitude and the requirement of orbiting above the equator result in high access latency for consumers, which renders satellite Internet infeasible for many use cases.[36,37]

The development of mobile communication has motivated improvements in the communication performance of satellites. The Iridium constellation was launched to provide voice and message service.[38] It consists of 66 active satellites in orbit at a height of approximately 781 km and inclination of 86.4°. The Iridium constellation is remarkable in that its satellites can maintain stability relative to each other even though they are in different orbital planes, and lower orbit altitudes require less radiated power for users and reduce the propagation delay.[39]

However, one major drawback of the Iridium constellation is its uneven coverage density. The Walker satellite constellation[40] can realize globally uniform coverage, but satellites from different orbital planes cannot maintain a static formation. The Walker constellation is defined by $i : h : t/p/f$, where $i$ is the inclination, $h$ is the altitude (the distance from the satellite to the ground, measured in kilometers), $t$ the total number of satellites, and $p$ the number of orbital planes. The phase parameter $f$ is the relative spacing between the satellites in adjacent planes. In particular, the Walker constellation is divided into two types: star and delta. The former uses near-polar orbits at an inclination of about 90 degrees. The latter uses orbits at an inclination generally less than 60 degrees, with the ascending node of the plane distributed over the full range of 360 degrees, and the satellites spaced evenly on the orbit. Emerging massive constellations such as Starlink and Kuiper are Walker delta constellations. An example of a $75^\circ : 800 : 40/5/1$ Walker $\delta$ constellation is shown in Fig. 2.

### 2.2.2. Satellite communications

Frequency is a crucial factor for satellite communications; microwave (300 MHz to 300 GHz) is suited to satellites because of its wide bandwidth and high frequencies, as well as the ability to use small antennas. Table 3 shows the frequency bands used by satellites. Most communication satellites work in the C frequency band, with an uplink frequency of 5.925 to 6.425 GHz and a downlink frequency of 3.7 to 4.2 GHz. At present, the X, Ku, and Ka frequency bands have been developed for civil service and broadcasting. However, the Radio Frequency (RF) bands, which are scarce, cannot keep up with the rapid increase of satellite applications. Higher frequency bands, such as the V frequency, even up to the optical frequency range (190 THz to 560 THz), are being explored. For example, Starlink[41] and Micius Satellite[42] are equipped with a laser communication payload.

Transmission losses in satellite communication links include free-space path loss, antenna misalignment loss, and polarization loss. In ground-to-satellite communication links, atmospheric absorption loss, and path loss in urban environments[43] also need to be considered. In addition to these factors, Doppler frequency also poses a significant risk to the efficacy of a satellite communication system. Among these, the free-space path loss $L_p$ accounts for the vast majority of the total transmission loss and is the primary loss being considered. It is given by

$$L_p = \left(\frac{4\pi d}{\lambda}\right)^2 \tag{1}$$

where $d$ is the distance between the satellite and the GS and $\lambda$ is the wavelength of the signal.

Doppler frequency shifts $f_D$ refers to the change in frequency or wavelength of a wave in relation to a satellite moving at high speed:

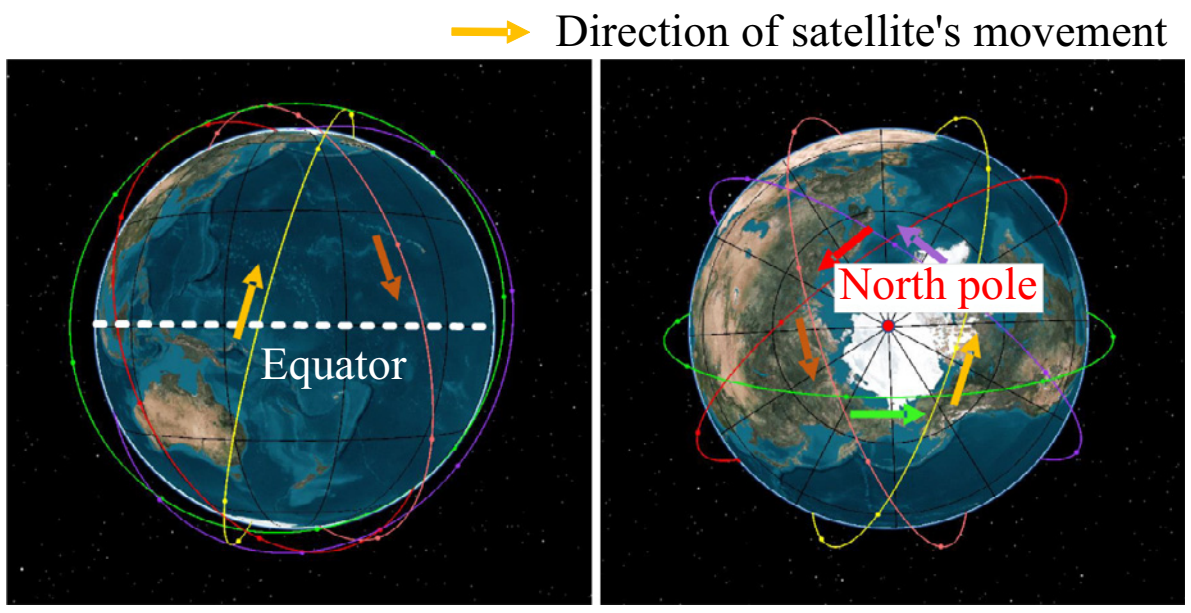

**Fig. 2** Illustration of a $75^\circ : 800 : 40/5/1$ Walker $\delta$ constellation.

**Table 3** Frequency bands used by satellites and their code.

| Code | Frequency band | Used by satellites (down-link/up-link) |
|---|---|---|
| UHF | 300 MHz to 3 GHz | 400/200 MHz |
| L | 1 GHz to 2 GHz | 1.6/1.5 GHz |
| S | 2 GHz to 4 GHz | 2.1/2.2 GHz |
| C | 4 GHz to 8 GHz | 6/4 GHz |
| X | 4 GHz to 8 GHz | 8/7 GHz |
| Ku | 12 GHz to 18 GHz | 14/12 GHz, 14/11 GHz |
| Ka | 27 GHz to 40 GHz | 30/20 GHz |
| V | 40 GHz to 75 GHz | 50/40 GHz |

$$f_D = f_c \frac{v_D}{c} \cos\theta \tag{2}$$

where $f_c$ is the carrier frequency, $c$ is the speed of light, $\theta$ is the angle between the connection of the satellite and the user, and the speed in the direction of relative motion is $v_D$. The "user" can be a GS, personal device, or another satellite.

Other losses include the antenna misalignment loss $L_{\mathrm{aml}}$, which arises from imperfections in the alignment of the transmitter's antenna with the receiver's antenna. The polarization loss $L_p$ is due to polarization mismatch, which arises from the misalignment between the polarization of the incoming wave and that of the receiving antenna subsystem. The atmospheric absorption loss $L_a$ arises from the interactions of free electrons and ions in the ionosphere, in addition to the absorption and scattering of electromagnetic waves by constituents such as oxygen, water vapor, and precipitation within the troposphere. The path loss in urban environments $L_u$ is due to the propagation of electromagnetic waves at low altitudes, shadow fading caused by buildings and mountains, as well as multipath effects.[43]

The overall loss can be represented as

$$L = L_p L_{\mathrm{aml}} L_p L_a L_u \tag{3}$$

and the received signal power $P_r$ can be represented as

$$P_r = P_t G_t G_r L \tag{4}$$

where $P_t$ is the transmitted power, while $G_t$ and $G_r$ are the gains of the transmitter and receiver antennas.

Within satellite edge computing, inter-satellite communication must be considered for computational offloading, federated learning, and resource allocation in satellite missions. Matthiesen et al.[44] divide Inter-Satellite Links (ISLs) into intra-orbit and inter-orbit links. Intra-orbit links connect satellites in the same orbital plane and form a stable ring network. Satellites in one ring network keep a relatively stationary position with each other. Instead, inter-orbit links connect satellites across orbital planes. In the Walker constellation, inter-orbit links are time-variant: Satellites can only remain linked for a short time.

### *2.2.3. Satellite network topology modeling*

A satellite network is a dynamic network composed of nodes and links. In this network, the nodes are the satellites, and the links are the communication links between nodes. Due to the time-variant positions of the satellite, the network topology is changing continually. In the Walker constellation, the links between satellites can only survive about ten or twenty minutes (depending on the satellite's orbit). In this subsection, we summarize the algorithm for modeling the topology of a satellite network, as in Ref. 45.

Assume the links in the network will be broken only if the two satellites are out of sight. The node can be modeled as $S = \{N, P\}$, where $N$ is the identifier of the satellite and $P$ is the position in space. The description of the position could be by either the orbit or the coordinates. The model of link $L = \{N_{S1}, N_{S2}, t_s, t_e\}$ indicates the two satellites $N_{S1}, N_{S2}$ and the times $t_s, t_e$ when the link is built and broken. Naturally, the difference $t_l = t_e - t_s$ represents the survival time of the link.

In fact, the change in the topology of the satellite communication network occurs at some specific moment. Therefore, the time of simulation model could be discrete. In Ref. 46, the topology is fixed during a period which is called a timeslice $T_i = [T_{si}, T_{ei}](i = 1, 2, \ldots)$, where $T_i$ is the $i$-th timeslice in the simulation time and $T_{si}$ is the time at which it starts and $T_{ei}$ is the time it ends. Naturally, $T_{s(i+1)} = T_{ei}$. Based on this, the time-variant topology can be modeled by $G_d = \{T, \{S_1, S_2, \ldots, S_M\}, \{T_1, T_2, \ldots\}, \{G_{s1}, G_{s2}, \ldots\}\}$, where $T$ is the duration of the simulation time. $\{S_1, S_2, \ldots, S_M\}$ is the set of satellite nodes and $M$ is their total number. In the $i$-th timeslice, $G_{si}(i = 1, 2, \ldots)$ describes the satellites (nodes) and links by $G_{si} = \{T_i, \{S_{i1}, S_{i2}, \ldots, S_{im}\}, \{L_{i1}, L_{i2}, \ldots, L_{in}\}\}$. This indicates that in the $i$-th timeslice, there are $m$ satellites, which have built $n$ links. The nodes are $\{S_{i1}, S_{i2}, \ldots, S_{im}\}$ and the links are $\{L_{i1}, L_{i2}, \ldots, L_{in}\}$. Based on this model, the network topology is a dynamic graph, and some OEC algorithms, such as computation offloading, mobility management, and resource allocation, can be simulated. Furthermore, the weights of the links in this model can be defined according to particular needs.

## *2.3. Mobile computing for IoT: From ground to space*

Over the last decade, the proliferation of mobile devices has facilitated remarkable progress in IoT and 5G/6G communication. The notion of the IoT, which refers to the ability of computers to sense information without human intervention, has been extensively applied to domains such as healthcare, environmental monitoring, home automation, and transportation.[47,28] With the burgeoning number of smart terminals, a massive volume of data is generated incessantly.[48] Fortunately, sophisticated communication technologies, with optical fiber and wireless communication, enable cloud-based computing and storage services at a remote data center.

**Table 4** Comparison of OEC and MEC.

| Feature | OEC | MEC |
|---|---|---|
| Edge server | Satellites | Base stations (BSs), wireless Access Points (APs), smartphones, etc.[8] |
| Hardware | Space-qualified chip (are evolving towards COTS now) | Mature COTS |
| Network | Satellite-GS or inter-satellite links with microwave, terahertz, laser communication technology. The delay in satellite communication ranges from a few to a few hundred milliseconds. | Internet with wireless communication and optical-fiber communication, with delays of tens of milliseconds or even microseconds and abundant bandwidth resources. However, long-distance transmission has extremely complex routing forwarding and electro-optical conversion. |
| Control and management | (1) The movements of satellites are predictable. (2) Schedule for satellite attitude adjustment, communication link establishment, and computation offloading. (3) Resource management cares about the energy consumed by the communications. | (1) The spatial locations of mobile devices are random so that the devices that can participate in MEC servers are randomly changing. (2) Schedule for computation offloading. (3) Resource management cares less about the energy consumption of communications. |
| Applications | Large-scale, on-orbit tasks such as automatic border monitoring, spatial situation awareness, and remote sensing image preprocessing, as well as providing access to computing for remote areas. | Latency-critical and computation-intensive applications such as AR/VR, autonomous driving and interactive online gaming. |

Nowadays, some extreme applications, such as AR/VR, Ultra-HD (UHD) video streaming, and autonomous driving, require extremely low latency or enormous computational ability, even both, which could be a heavy load for network and cloud services. Cloud computing will not be efficient enough to support these applications.[30] Hence, in 2014, mobile edge computing was first proposed by the European Telecommunications Standard Institute to obtain a lower latency.[49]

However, due to its limited coverage, the terrestrial network cannot provide seamless access, especially in remote areas (such as deserts, forests, oceans, etc.) and disaster areas. In this context, the satellite communication network is a vital supplement for terrestrial users. With the cost of satellite manufacturing and launching reduced, the space-air-ground integrated network is expected to play a pivotal role in the IoT.[2] To address the limitations of the bent-pipe architecture, the OEC was first proposed in Ref. 29. By organizing a set of nanosatellites that orbit on the same path (i.e., flying in formation) into pipelines that parallelize both data collection and data processing, OEC can reduce the ground infrastructure and system edge processing latency compared to a bent-pipe architecture. Similarly,[17] describes power and software optimizations for the OEC based on this formation flying architecture, verifying the feasibility of distributed computing in space. After the concept of space distributed computing was proposed, inter-satellite communication and on-board processing capability became the bottlenecks.

In addition, edge computing can be implemented on a virtualized platform, allowing high-bandwidth and low-latency applications while decreasing privacy and security risks as well.[50,51] Moreover, OEC has been verified to support more advanced computing in space.[29]

## 2.4. *Orbital edge computing versus terrestrial mobile edge computing*

As shown in Table 4, there exist significant disparities between OEC and MEC systems in terms of architecture, management, and applications. Compared with MEC, OEC has the advantage of a more extensive and even coverage. More details are in the following.

### 2.4.1. *Architecture*

The definition of MEC refers to the use of edge servers (such as BSs) for offloading computational tasks from mobile devices. Due to being closer to the data source, it has less latency than Cloud computing. Inspired by this, the satellite undertakes some computing tasks in orbit through OEC, rather than transmitting all the raw data back to the ground. In OEC, some satellites equipped with computing hardware provide computing resources to users.

### 2.4.2. *Computing*

The software, hardware, and systems on the satellites are completely different from those on the ground computing devices in MEC: they are constrained by the volume and weight of the satellites, as well as by the harsh environment of space.

Computing hardware in smartphones consists of a System on Chip (SoC), such as the A15 chip in iPhone14, which has six cores based on a 5-nm technology node and provides 8.6 W of power and 15.8 Tera Operations Per Second (TOPS).[52] Unlike MEC, satellite hardware must be more reliable due to the challenging space conditions. The first smart satellite funded by the European Space Agency, called Φ-Sat, carried the Movidius Myriad 2 Video Processing Unit (VPU)[53] based on 28-nm technology. This VPU, with only 0.98 W of power and 2 TOPS, was used for hyperspectral images for cloud detection based on a Convolutional Neural Network (CNN).[54]

### 2.4.3. *Control and management*

The advantage of satellite communication is the ubiquitous connection and lower latency compared to terrestrial networks over long distances.[19] For example, the Round-Rrip Time (RTT) for Internet communication between New York and London is about 76 ms. On the other hand, the minimum RTT achievable via optical fiber that follows a great circle path is 55 ms. However, the use of ISLs can further reduce the RTT to only 50 ms.[55]

However, establishing a communications link (especially laser communication) on a satellite requires several minutes of prior attitude adjustment for antenna alignment, which differs from terrestrial wireless communication. Therefore, before computation offloading, an OEC manager needs to design instructions for task scheduling according to the satellite's motion, unlike for ground networks.

The energy consumption for communication is often overlooked in MEC. However, it is more of a concern in OEC, as the distance traveled by the electromagnetic waves for satellite communication often exceeds hundreds of kilometers, which consumes more energy.

#### *2.4.4. Mobility*

MEC is troubled by the random movement of ground users, such as AR, autonomous driving, and other applications.[8] Therefore, it faces challenges in realizing ubiquitous and reliable service connections. Relative to the speed of the ground users, the challenge for satellites is their own high-speed movement and unstable connection, so the orbital characteristics of satellites produce completely different challenges for OEC. Therefore, MEC focuses on user mobility awareness,[56] while OEC pays more attention to orbit prediction and multi-satellite collaboration in mobility management.[57]

#### *2.4.5. Application*

OEC is an important supplement to MEC, providing access to computational services for users in remote areas and disaster areas, which the terrestrial Internet cannot cover.[19] OEC also offers computational offloading services for in-orbit spacecraft, such as in-orbit processing of remote sensing images, etc.[58] In addition, OEC can boost the ocean and low-altitude economy.[14]

### *2.5. Advanced technology in space*

In the following subsection, we introduce recent advances in technology that could be beneficial to the organization of OEC, including laser communication, smart satellites, and space-air-ground integrated networks.

#### *2.5.1. Laser communication*

Free-Space Optical (FSO) communication in outer space offers the potential for high-speed, high-capacity data transmission using compact and energy-efficient onboard equipment.[59] The narrower beams of FSO communication links can reduce interference and enhance security.[60] As a result, it serves as a precursor to the next-generation technology, laser communication, which aims to enhance both communication rates and ranging accuracy within satellite networks. Laser ISLs offer significant advantages over traditional microwave communication, including large bandwidth, high data rate, ease and speed of deployability, and low power. The first laser ISLs were demonstrated in 2008 between Terra SAR-X and NFire, two LEO satellites located 5 500 km apart and moving at a speed of 25 000 km/h, achieving a capacity of 5.5 Gbps. Studies have shown that laser communication technologies are superior to RF technologies in terms of data rate, security of the communication, and weight reduction.[57,59,61] Table 5 provides a summary comparison of laser and RF communication. One of the most promising applications of laser communication is SpaceX's Starlink project, which plans to use inter-satellite laser communication technology to provide Internet services up to 1 Gbps with the deployment of 12 000 satellites.[41]

Although laser communication provides high-speed links, FSO communication still suffers performance degradation due to the following:[61]

(1) Imperfect propagation path, various gases and aerosols in the atmosphere induce absorption and scattering losses of light.
(2) Atmospheric turbulence, induced by fluctuations in temperature and pressure within the atmosphere along the propagation path, influences the trajectory of light propagation.
(3) Background noise, which comes from the Sun and other stars.
(4) Pointing loss, caused by satellite vibration or imperfect tracking-and-stabilization mechanisms, reduces the signal strength. FSO communication links have lower beam spread and higher directivity, which pose significant challenges to their alignment with the desired receiving station.[62]

Recently, mixed RF/FSO communication networks have been proposed to alleviate these shortcomings of FSO.[63,64] Compared to conventional RF communication systems, mixed RF/FSO communication significantly improves data transmission rates, throughput, and real-time reception performance.[65] OEC faces the challenge of high-frequency data communication.[66] The development of OEC urgently requires communication methods with larger bandwidth and faster speed. Research on FSO communication is addressing this.

**Table 5** Laser versus RF for ISLs.

| Characteristic | Laser ISLs | RF links |
|---|---|---|
| Spectrum | Near infrared and visible light band | Ka, Ku and mm-wave band |
| Bandwidth | In THz | In GHz |
| Wavelength | In nm | Thousands of times larger than the optical wavelength |
| Beam spread | Much narrower | Typically 1 000 times more |
| Power consumption | More received power for a given transmitted power | Less received power for a given transmitted power |
| Security | Very difficult to intercept due to high directivity | Easy to intercept |
| Hardware | Less volume and less weight | More volume and more weight |

### 2.5.2. Massive MIMO

Massive Multiple Input Multiple Output (MIMO) is a technology that uses a large number of antennas at the transmitter and receiver to improve communication performance. It promises to meet the ultra-high data rate requirements and high robustness of 6G, making it a key technology for future satellite communication systems.[67]

The rapid development of mega-constellations has brought new opportunities for applications of massive MIMO in satellite communications. You et al.[68] introduced massive MIMO into LEO satellite communication systems, and simulation results demonstrate that it significantly enhances the data rate of LEO satellite communication systems. Abdelsadek et al.[69] proposed a new LEO satellite network architecture based on distributed massive MIMO technology, allowing ground user terminals to be connected to a cluster of satellites.

Massive MIMO enables simultaneous data transmission to multiple users within the same frequency band, effectively increasing spectral efficiency. Massive MIMO allows for highly focused beamforming, concentrating signal energy towards specific users, reducing interference, and improving Signal-to-Noise Ratio (SNR), which leads to higher data rates.[68]

However, as an emerging technology, Massive MIMO still faces several challenges:

(1) Dynamic topology: The dynamic network topology of satellites entails issues with defining the cluster that serves the user terminals (i.e., the number of satellite access points and which satellite access points serve each user terminal).[68]
(2) Synchronization: Cooperative transmission across massive MIMO between a cluster of satellites, synchronization in time, frequency, and phase is required. This requires significant computational ability for data processing to achieve equivalent time synchronization, or the design of a complex control system to dynamically adjust the reference frequency of the satellite-based on time differences.[70]
(3) Overhead: The use of machine learning techniques for massive MIMO in channel estimation results in increased computational complexity, which further burdens the computational resources of satellites.[71]

Massive MIMO is a crucial technology for the development of OEC. It is capable of effectively supporting high-density data transmission and real-time computing requirements, providing strong technical support for future intelligent and networked applications.

### 2.5.3. Smart satellite

A satellite is typically developed for a single or a small number of specific missions, which determine its capabilities, functions, and operations. The hardware and software on a satellite are tightly coupled, making the whole system inflexible and non-universal. However, certain architectural components, such as communication systems and antennas, can be shared between different missions. The Software-Defined Satellite (SDS) has been proposed as a potential solution to these challenges, improving the efficiency of a satellite network by using software-defined radio as its universal hardware platform and defining its functions via software.[72] The Chinese Academy of Sciences (CAS) has developed and launched the experimental Tianzhi series of SDSs. The Tianzhi No. 2 D-star is capable of supporting super-resolution image reconstruction on the satellite due to its 40-TOPS computational power and only 19 kg weight. The successful application of SDS has facilitated the development of smart satellites. However, conventional task scheduling requires GSs to pre-arrange instructions, which are uploaded when satellites come into range. As satellite tasks become more complex, this method is inefficient. In the future, autonomous task scheduling could alleviate these limitations, enabling satellites to make decisions online with minimal manual intervention. Advances in smart satellite research, such as the creation of satellite computing networks[73] and space-air-ground integrated networks, will stimulate numerous OEC applications. Especially, a constellation of computational nanosatellites presents novel research challenges and opportunities in distributed sensing, computing, and communication.[33]

Smart satellites are continually evolving and increasingly exhibiting the following characteristics:

(1) Small satellites. Mass-producible small satellites reduce the cost of both manufacturing and launching, which facilitates the development of mega-constellations.[74]
(2) Efficient energy utilization. This is difficult for small satellites since their small size provides little room for energy storage. So they need efficient resource management algorithms to plan the use of on-board energy.[33]
(3) Hybrid-processor architectures. Computing payloads in satellites integrate multiple computing architectures to gain their collective advantages, such as a combination of CPUs, GPUs, FPGAs, and DSPs.[10]
(4) Generalized operating system. Satellite components are becoming more generalized, and they need to support general operating systems such as Linux to facilitate the development of various on-board software, in order to meet the rapidly growing demands of business.[11]
(5) Lightweight virtualized resource management. Virtualized resource management tools, like Docker[1], can effectively manage and orchestrate node resources, enhancing the flexibility of deploying on-board applications.[34]

Space-Air-Ground Integrated Network (SAGIN), a network architecture that interconnects space, air, and ground segments has attracted attention in order to meet the demand for seamless connectivity by the future digital society.[2,26] High Altitude Platform stations (HAPs) and Unmanned Aerial Vehicles (UAVs) can provide IoT services in infrastructure-less environments to help users conveniently access the IoT services without the construction of an expensive infrastructure. Fig. 3 shows the three primary segments of SAGIN: space, air, and ground; satellites serve as edge nodes for the space section, and UAVs, aircraft, and other HAPs constitute the air section.

OEC presents a vision of ubiquitous computing service accessibility within SAGIN. Yu et al.[75] proposed that OEC can support the Internet of Vehicles (IoVs) in remote areas. Similarly, Cheng et al.[76] proposed a task offloading framework through resource slicing in SAGIN for vehicles. However, the communication and computing capacities accessible via SAGIN are presently constrained, particularly those deriving from aerial or satellite networks. In SAGIN, Software-Defined Network (SDN) and Network Functions Virtualization (NFV) are crucial technologies for resource orchestration.[77] These technologies have motivated the use of network slicing and virtualization in satellite networks, particularly in OEC scenarios. Such strategies facilitate the optimal utilization of energy and computational resources. Kim et al.[78] explored feasible orbit heights of IoT-supported satellite constellations, and proposed scheduling rules for sliced satellite networks and appropriate offloading strategies for OEC. Giuseppe et al.[79] applied device virtualization to CubeSats constellations to orchestrate the satellites so as to provide efficient OEC se rvices on the ground.

[1] https://www.docker.com

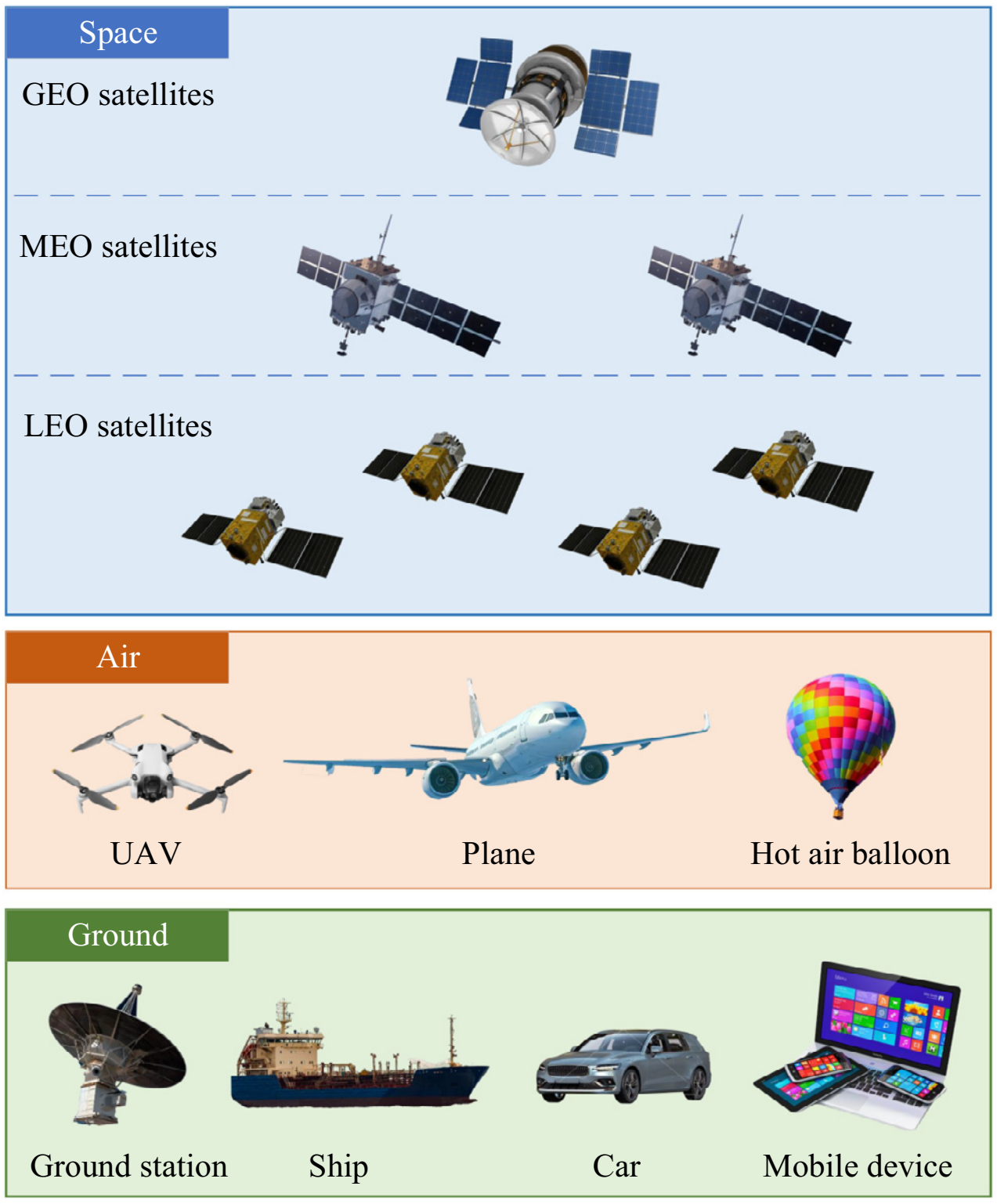

**Fig. 3** The architecture of a space-air-ground integrated network consists of three network components that can communicate with one another via RF signals. The space network includes satellite constellations categorized as GEO, MEO, and LEO. The air network includes UAVs, aircraft, and balloons and is an aerial mobile system. The ground network, also known as the terrestrial network, is mainly made up of cellular networks, mobile ad hoc networks, and Wireless Local Area Networks (WLAN).

## 3. Systems

In this section, we introduce the OEC systems' satellite hardware, satellite software, and architecture. First, we introduce the satellite hardware platform that can be used in edge computing applications. Second, we introduce the satellite's software and Operating Systems (OSs). Lastly, we will introduce the architecture of satellite edge computing.

### *3.1. Satellite hardware*

Computers on satellites are often exposed to high-intensity radiation, usually from cosmic rays (mainly protons and $\alpha$ particles), solar-weather events (including solar wind, solar flares, and coronal mass ejections), and the Earth's magnetic field (mainly composed of electrons, protons, and some heavy ions).[80] Although satellite platforms incorporate radiation-shielding shells to mitigate the effects of these, the electronic components aboard the satellites might still interact with these particles. George et al.[10] categorized these interactions into two types: long-term cumulative effects and short-term transient effects. Cumulative effects include a buildup of Total Ionizing Dose (TID) levels, ionization of circuits, Enhanced Low-Dose-Rate Sensitivity (ELDRS), and Displacement-Damage Dose (DDD). The Single-Event Effects (SEEs) include Single-Event Upsets (SEUs), Single-Event Transients (SETs), Single-Event Latchups (SELs), Single-Event Burnouts (SEBs), Single-Event Functional Interrupts (SEFIs), and Single-Event Gate Ruptures (SEGRs). These have the potential to induce software malfunction or even irreversible hardware failure in satellites. As a countermeasure, an array of fault-tolerant computing techniques has been integrated into space systems, guaranteeing their reliable functioning in space. Traditional methods for achieving this fault tolerance depend on the redundancy of the hardware, information, and network. Hardware redundancy is accomplished by introducing additional components into the design, such as utilizing three or more processors to execute identical tasks, culminating in a majority vote for the final result. Information redundancy, on the other hand, is achieved through techniques like Error Detection And Correction coding (EDAC), Error Correction Code (ECC), Cyclic Redundancy Check (CRC), Algorithm-Based Fault Tolerance (ABFT), and parity checking. Network redundancy, meanwhile, necessitates redundant network connections and paths within the network topology.[10]

A considerable challenge for the OEC is the necessity for the satellites to provide high-performance and reliable computing services (goals which are often at odds with each other) using limited resources (power, size, weight, and cost). Traditional onboard computer hardware often uses expensive, low-performance components that lag behind the latest advances. On the other hand, Commercial Off-The-Shelf (COTS) computing hardware is more feasible for OEC, as it offers superior performance, and in key tasks such as flight control and satellite operations, edge computing services are superior.[81] These COTS normally have a hybrid architecture, which mixes dissimilar computing technologies (CPU, GPU, DSP, FP, etc.) to take advantage of their respective strengths.[82] George et al.[10] analyzed the vast performance advantages of SoC in hybrid computing for OEC from the perspectives of cost and radiation resistance.

Sheldon et al.[83] analyzed the radiation performance of the Snapdragon SA8155P SoC under a 1-mm aluminum shielding box at a height of 550 km on circular orbits of different inclinations, finding that a 1-mm aluminum shielding box is sufficient to protect the SoC for up to five years. Based on this, Pfandzelter et al.[81] estimated the soft error rate caused by SEUs to be $10^{-3}$ to $10^{-4}$ per day per device and estimated that with a similar computing resource equipped in the Gen1 Starlink satellite constellation (60 SoCs in each of the 4 408 satellites), 30 to 300 errors induced by SEUs will occur every 24 h.

Φ-Sat-1, equipped with the Intel Movidius Myriad 2 hardware accelerator, is the first satellite to employ an onboard deep neural network for satellite earth observation. It has proved to be robust against ionizing radiation, developing a Cloudscout segmentation neural network, run on Myriad 2, to identify, classify, and eventually discard on-board cloudy images.[84]

Bhattacherjee et al.[85] carried out a feasibility study of on-orbit computations by leveraging a high-end server (ProLiant DL325 Gen10 developed by Hewlett Packard Enterprise[2]) with 64 CPU cores, a primary frequency of 2.4–3.35 GHz, a maximum capacity of 2 TB memory, and a weight of 15.6 kg. Wang et al.[9] deployed the Atlas 200 DK[3] on an in-orbit satellite belonging to the Tiansuan constellation. This satellite affords a maximum of 22 TOPS of CPU computational resources, consuming only 20 W.

Recently, Xu et al.[86] proposed a novel server architecture consisting of massive low-power System-On-Chips (SoCs). A quantitative analysis showed this server's superiority over conventional servers (Intel CPU, NVIDIA GPU, etc.) in terms of three critical in-space computing metrics: energy efficiency, weight, and volume. This inspired OEC researchers with a hardware platform that can be applied to OEC and provided guidance for simulating OEC on actual facilities.

More and more COTS hardware is being applied in satellites, providing computing support for the OEC. At the same time, we also call for more research on OEC hardware. This includes radiation resistance, improving the performance of hybrid computing hardware, and more real on-orbit demonstrations.

### 3.2. Satellite software

Traditionally, onboard software has been written in something close to hardware in assembly language, Ada, C, or C++.[87] The OS provides task and resource management and application software for providing mission-specific functionality. An OS employed on terrestrial computers and mobile devices would be unsuitable for direct application within satellites due to the harsh environment in space. For traditional satellite services, reliability and stability are the primary considerations in satellite system design.

Satellite hardware and software have recently been evolving at an accelerated pace, with a trend towards generalization. However, traditional satellite systems struggle to match the emerging trends in generalized software and hardware.

In recent years, there has been an increasing utilization of Linux in spacecraft on-board software, including Planet and SpaceX.[88] However, the Linux used in spacecraft differs from that used in Personal Computers (PCs) or embedded devices. SpaceX has developed a self-made, deeply customized, distributed Linux for its spacecraft, which is not a third-party distribution. Additionally, it includes drivers designed specifically for dedicated hardware, as well as security enhancements at the hardware level.

Satellite IoT urgently needs to develop a generalized operating system that can be loaded onto satellites.[28] Wang et al.[89] discussed the possibility of dynamically loading software in orbit and the potential of providing services coordinated with the Cloud. To enable satellites to evolve into space edge computing nodes, the authors also aimed to standardize hardware platforms and OSs. This solution can also facilitate the orchestration and management of resources in a satellite network.

The next generation of satellite OS is gradually revealing the following characteristics:

(1) Compatibility: It should be compatible with the existing Linux software ecosystem, allowing smooth migration of existing software to satellites.[11]
(2) Upgradeability: Generalized OSs and varied software applications require that satellites promptly update their systems and software to address system vulnerabilities, software inaccuracies, hardware deterioration, etc.[90]
(3) AI-supported: Machine learning or deep learning in space is an important direction for OEC moving forward. Satellite OS should be AI-algorithm friendly.[91]

### 3.3. Architecture

Fig. 4 shows the architecture of OEC and how it provides edge computing services. In this architecture, we have delineated the roles assumed by GS, LEO, and GEO satellite constellations. In the following subsections, we describe this architecture from the perspective of the cloud-edge-end three-layer organization framework and discuss how OEC provides services to both airborne and ground users.

#### 3.3.1. Cloud

GS and some MEO and GEO satellites can respectively play the roles of a ground-based cloud and a space-based cloud.[20] Their functions include resource orchestration, mobility management, access control, task scheduling, etc. The difference is that in the ground-based cloud there are abundant computing and storage resources, while the space-based cloud only has limited resources (but much more than edge satellites). Therefore, the ground-based cloud can have a global view, coordinate and control all in-orbit satellites, and have enough computing ability to predict the global satellite constellation network topology. In addition, GS can access the terrestrial Internet and send computing tasks to data centers. Data centers have the most computing and storage resources and can handle user-generated computing tasks.[73] GS also collects telemetry data from satellites to monitor their operating status. Space-based cloud satellites can orchestrate low-orbit satellites within a specific coverage area, thereby assuming some of the responsibilities of the terrestrial cloud centers, such as the regulation and supervision of satellites in their coverage.[92] This approach largely eliminates their dependence on directives from GSs, enabling satellites to autonomously manage resources and schedule tasks through space-based intelligence.

#### 3.3.2. Edge

Hu et al.[93] divided the OEC architecture into four layers: hardware, virtualization, control, and service. The hardware layer has the payload of the satellite, including physical devices such as communication, sensors, computing, and storage. The virtualization layer abstracts physical devices into available resources through virtualization and container technology.[58] The service layer realizes the isolation of tasks and resources through Virtual Machines (VMs) running various applications. Xie et al.[73] described the services that OEC can provide and the function of the control layer. Due to the impact of changes in a satellite's orbit and other perturbations, the orbit information needs to be updated regularly. The control layer manages this information and regularly updates it to the Cloud, which ensures the accuracy of the Cloud's prediction of the network topology. The control layer is also responsible for the health management of the satellites, including updates and rollbacks of software and systems, monitoring software and hardware errors, etc. This information is regularly synchronized from the edge nodes to the cloud nodes. Cloud servers can schedule tasks globally and optimally by leveraging network topology, resource statuses, and node health.

[2] https://buy.hpe.com/hk/en/compute/rack-servers/proliant-dl300-servers/proliant-dl325-server/hpe-proliant-dl325-gen10-server/p/1010868976
[3] https://e.huawei.com/en/products/computing/ascend/atlas-200

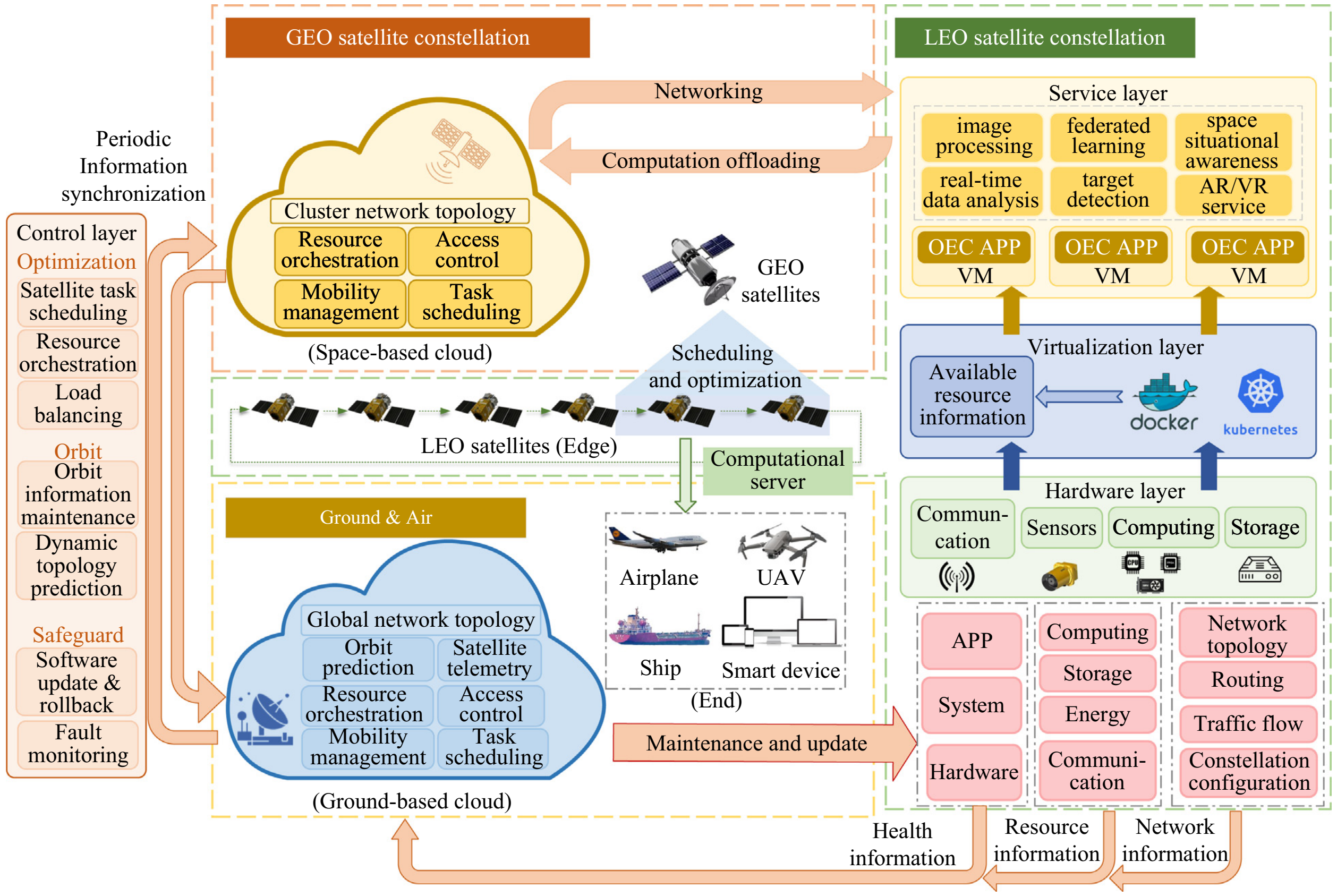

**Fig. 4** Architecture of OEC systems. (A) "GEO satellite constellation" plays the role of "Space-based Cloud," offering functionalities such as resource scheduling, network management, and maintaining orbit information, but only managing a limited range of satellite clusters and networking them. LEO satellites can offload computing tasks to these GEO satellites. (B) "LEO satellite constellation" is the "Edge", which can offer edge computing service to "End". They periodically report their status, e.g., available resources, health condition, and network to the "Cloud" for task scheduling. (C) GS serve as "Ground-based Clouds" and have abundant computational and storage resources. They routinely synchronize information with GEO satellites to ascertain the global status of all satellites. (D) The user end, such as airplane, UAV, ship, and mobile phone, which is the originator of the service. Computational tasks and data are generated here.

### *3.3.3. End*

In the OEC architecture, the entity being serviced is referred to as the "end," acting as the origin of the computational tasks or data. When processing on-orbit images, remote sensing satellites (acting as an "end") can offload inference tasks to the "edge." On the ground, the OEC can offer computational access to remote, marine, and disaster-stricken areas[94] or facilitate access to terrestrial edge computing for the benefit of urban vehicles and intelligent devices.[95] In the air, OEC can extend its services to UAVs and airplanes.[96] Zhang et al.[97] proposed a three-layer aerial edge computing network in which satellites receive task data from UAV and ground units. Through in-orbit data analysis and processing, they plan trajectories and schedule tasks for UAV clusters, providing IoT services for the coverage area.

## 4. Applications

### *4.1. Cases of OEC applications*

An increasing number of emerging applications can benefit from MEC by offloading their computation-intensive tasks to servers located nearby. Achieving lower task latency while fully utilizing edge devices' computing resources. The potential use of this technology in space is also great. In this section, we present two examples to demonstrate the paradigm of OEC.

OEC can provide services for users on the ground. Li et al.[16] developed a prototype to verify the benefits of OEC. As shown in Fig. 5, they uploaded the original video to the LEO satellite network and downloaded the processed video. Compared with transmitting data to cloud servers on the ground through a satellite network, using the OEC had lower service latency and occupied less bandwidth. More generally, users offload the computing task to an LEO constellation. Then, the OEC controller can determine whether the task should be offloaded to the Cloud or executed in Edge because it has a global view. When users are in remote areas or some areas without access to a terrestrial network, the OEC can provide ubiquitous computing power access services. This provides very attractive applications in some scenarios. In situations where disaster-stricken areas are inaccessible to ground networks, OEC can use remote sensing satellites for in-orbit evaluation of the disaster. This implementation can help quickly establish a reliable and resilient network, which is essential for efficient disaster management. Furthermore, OEC can optimize the scheduling process of UAV clusters through intelligent algorithms.

OEC can provide numerous applications in space, such as task scheduling and offloading. When satellites are in the same orbital plane, they are referred to as flying in formation[98] and are relatively static to each other. This implies that they can establish stable communication links all the time. Therefore, reasonable task scheduling will avoid crossing orbits. Nowadays, reinforcement deep learning has been shown to be very good at task scheduling and offloading.[99] As Fig. 6 shows, when satellite $A$ generates a change detection task, it decides whether to offload to another satellite. If offloading, under the guidance of GEO satellites, some free satellites (such as $C$) will be allocated for $A$'s edge computing servers. Then, $A$ will divide the data into several slices and offload them. OEC has improved the efficiency of remote sensing applications. Traditionally, under the bent-pipe architecture, remote sensing images are sent to the GS for processing. A remote sensing satellite gathers 10 TB each day, and a weather satellite 760 GB. OEC is expected to mitigate this intense bandwidth consumption.[34] Wang et al.[11] have verified on a real satellite that OEC can reduce the image detection latency from days to minutes without losing accuracy.

Currently, existing applications include deep learning-based low-light satellite image enhancement in orbit,[100] 6G based GS-satellite collaborative edge computing services,[101] and in-orbit encoders for semantic communication.[102]

OEC has other practical applications in various scenarios, such as space intelligence, autonomous vehicles, and AR/VR. This section will discuss some of these applications and the benefits of using orbital edge computing. As shown in Fig. 7, we categorize these applications based on the location of the data source, with satellites serving as computing nodes and relay nodes.

### *4.2. Source of data is ground*

As shown in Fig. 7(a), satellites can serve as computing or relay nodes to process data received from the ground. The GS schedules the missions and sends instructions to the satellites. According to the instructions, the satellites adjust their attitude, align the antenna, transmit data, compute results, etc. Applications in this scenario are highly sensitive to latency. Terrestrial networks are unsatisfactory in remote transmission due to the complexity of the routing, whereas satellite communication offers advantages in terms of coverage and simplicity of the routing between remote areas. Satellite edge computing can adapt to applications that require a significant amount of computing resources.

#### *4.2.1. AR/VR and UHD video*

AR and VR displays have the potential to serve attractive applications, including health care, education, engineering design, manufacturing, retail, and entertainment,[103] and UHD video can provide better visual differentiation and better depth perception for users.[104] However, they face many challenges due to the complexity of the computations, including encoding/decoding and rendering, involving more processing time.[105] Edge computing provides opportunities for these applications by transferring part of the computing pressure to edge devices. However, terrestrial networks may fail to provide ubiquitous coverage to suburban and rural areas, and the latency obviously increases with the distance of the transmission. As an effective solution, OEC provides ubiquitous access and ultra-long distances with low-latency transmission, opening up further possibilities for these applications.[106]

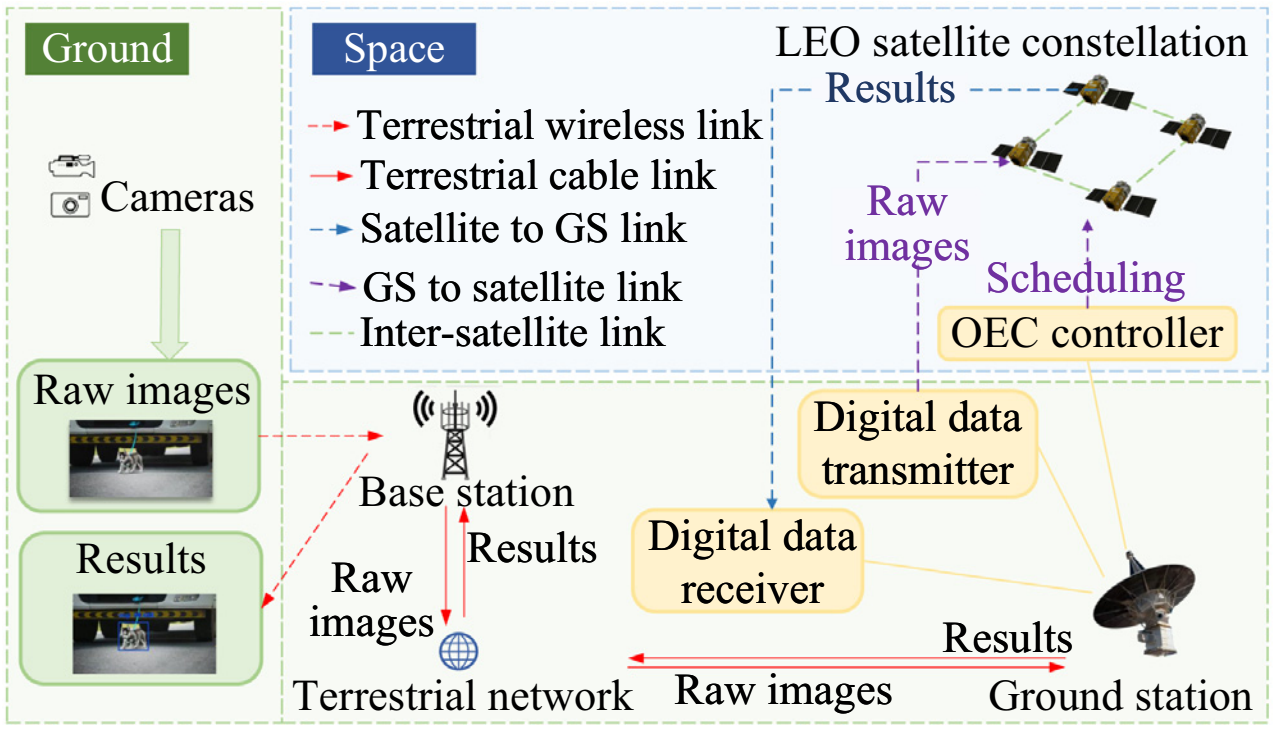

**Fig. 5** Framework of object tracking application in OEC.[16] Cameras on the ground can send images to the LEO satellite network. Then the processed images. are downloaded.

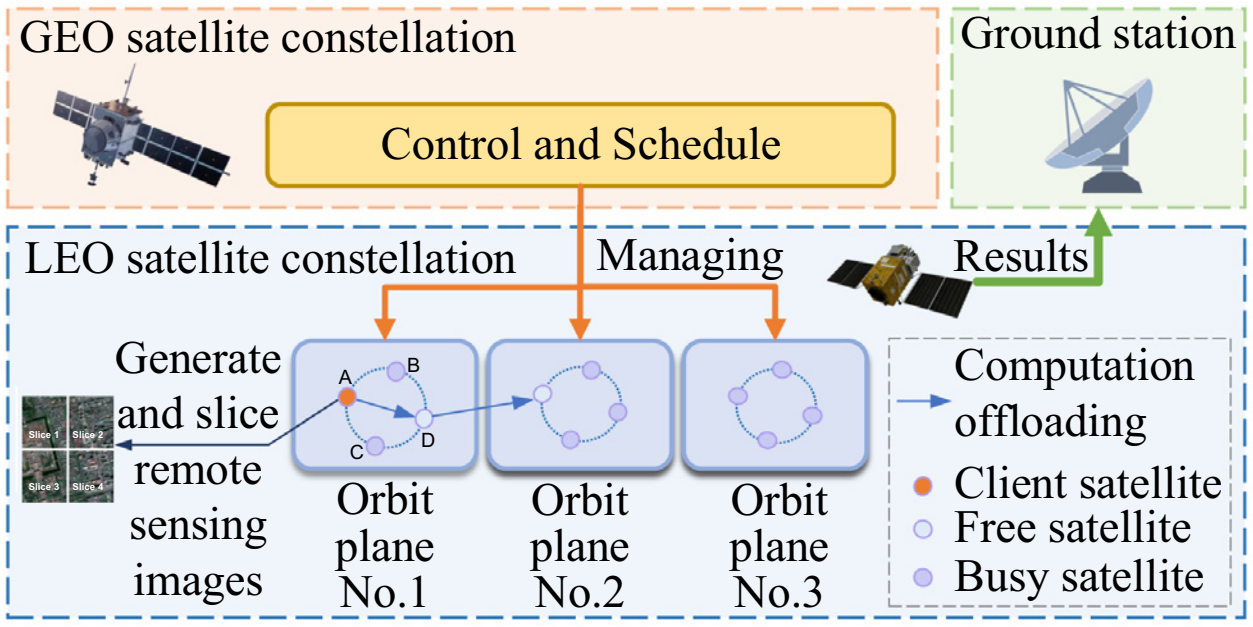

**Fig. 6** Framework of disaster surveillance in OEC.[34] LEO remote sensing satellite $A$ shot images and slices them for edge computing. GEO satellite guides $A$ to offload this task to free satellites (same or different orbital planes from $A$). Then the disaster evaluation can be done by OEC and the results can be sent. back to the GS.

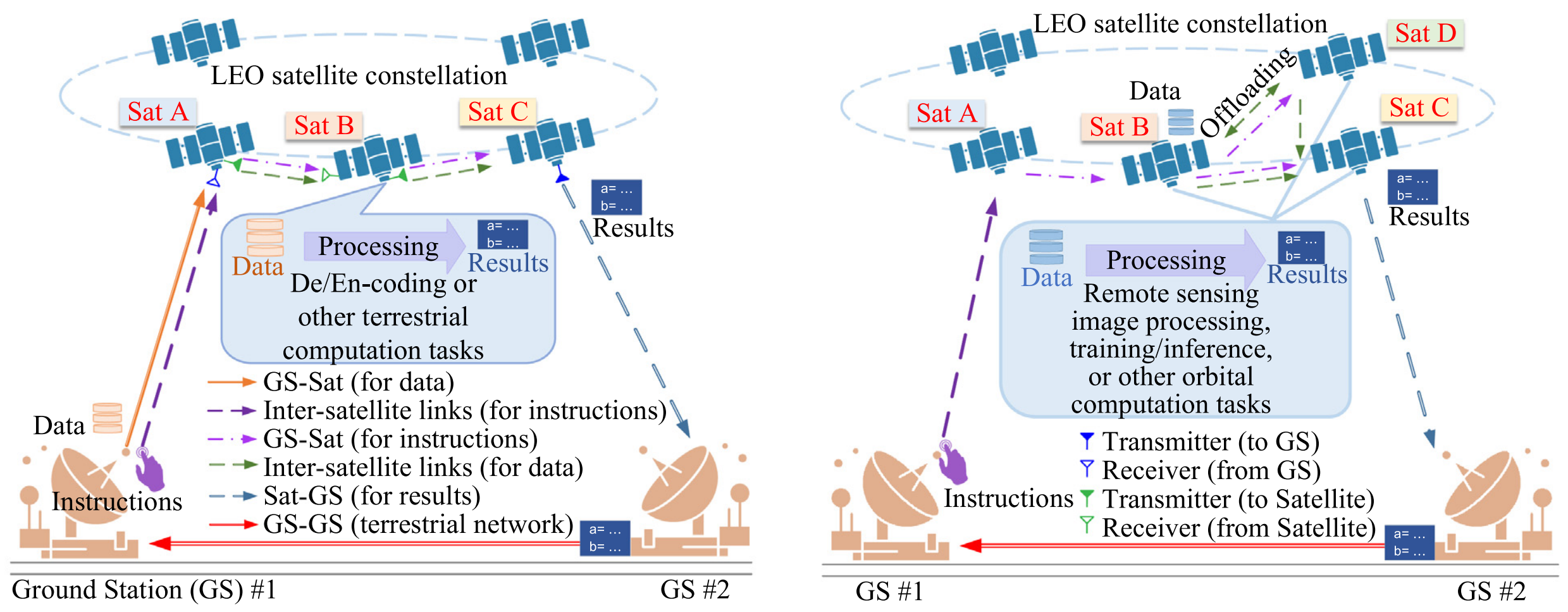

(a) Data source is at GS.

(b) Data source is at satellite.

**Fig. 7** Satellites served as computing and relay nodes. (a) Satellite (Sat A) accepts data and instructions from the GS #1 and shares them with Sat B and Sat C; data includes video/audio streaming or other terrestrial computation tasks, and the instructions include task scheduling, satellite attitude adjustment, etc. The data will be processed by a computing node (Sat B), and then the results will be sent to GS #2 via a relay node (Sat C). Finally, GS #1 can get the results from GS #2 by the terrestrial network. (b) The computing node (Sat B), which is also the data source, accepts instructions from GS #1 via a relay node (Sat A). It may have remote sensing images, data of scientific experiments, or other orbit data. Sat B can process them on orbit and GS#1 receives the results from GS #2 (terrestrial network) and a relay node (Sat B).

Celestial[19] evaluated the performance of WebRTC video meetings using OEC based on the Starlink network. Three users sent a high-definition video stream with 2.6 Mb/s while receiving video from the others. The experiment showed that the end-to-end latency of both the satellite servers and cloud servers for at least 80% of the duration of the video conference was below a maximum RTT of 16 ms and 46 ms, respectively. This confirms that satellite servers have the potential to significantly improve the quality of service for clients in latency-sensitive applications.

### *4.2.2. Internet of vehicles and tntelligent traffic*

The IoV is a combination of vehicle auxiliary devices and roadside infrastructure that aims to enhance the safety, security, and efficiency of roadway transportation. It also offers different services to various types of vehicles, including private cars, taxis, and buses. Satellites have become a vital element in providing positioning and navigation services for vehicles, and they are an essential component of IoV.[73,107] Yu et al.[75] suggest addressing the challenge of providing vehicular networking services in remote or disaster-stricken areas by using edge computing and satellite networks. Their proposed fine-grained joint offloading and caching scheme leverages the space-ground collaboration to provide communication services for vehicles in rural areas, isolated islands, and remote places where terrestrial infrastructure is not available. The scheme aims to provide real-time Enhanced Cellular Satellite Assisted Global Navigation System services for terrestrial vehicles.

## *4.3. Source of data is space*

As shown in Fig. 7(b), satellites can serve as a data source and participate in joint training/inference of models through inter-satellite links. Additionally, GSs can send models or instructions to the satellites. This paradigm offers a significant advantage by avoiding extensive data transmission between satellites and GSs. This is beneficial because it saves on satellite communication costs and improves the intelligent decision-making capabilities of satellites in orbit.

### *4.3.1. Earth observation*

Satellites are essential for capturing earth data in various fields, such as weather, vegetation, and urban development. The data generated is growing rapidly in terms of size and variety, which creates immense pressure on space-ground communication.[108] However, OEC can analyze this data, provide near real-time analysis, and significantly enhance decision-making. One specific application of this technology is the real-time monitoring of natural disasters. By quickly detecting and transmitting critical information to GSs, response times can be greatly improved.[29] Satellite edge computing can also identify invalid images caused by cloud cover, which reduces the cost of data transmission to the ground.[9] Environmental monitoring is another potential application of satellite edge computing. Tracking changes in land use, forest cover, and ocean temperature provides an opportunity to quickly identify changes and send alerts to stakeholders for timely action. Moreover, multiple agile satellites can utilize satellite edge computing to solve the scheduling problem, enabling on-orbit decision-making and intelligent mission planning.[109] These applications demonstrate the potential of satellite edge computing to enhance the efficiency and effectiveness of earth observation missions.

### *4.3.2. Scientific research*

Scientific Experiments satellites such as DArk Matter Particle Explorer (DAMPE), Quantum Experiments at Space Scale (QUESS), etc. are used to gather data for scientific research. Scientists obtain a large amount of experimental data from these satellites for space exploration and other missions, which often require transmission via satellite–ground links for subsequent analysis on the ground, a process that is inefficient. Edge computing on satellites, however, seeks to minimize data transmission and even enable in-orbit data analysis and processing.

Several satellites devoted to scientific experiments, including the Dark Matter Particle Explorer (DAMPE) and Quantum Experiments at Space Scale (QUESS), are instrumental in gathering data for space exploration and other scientific research missions. Although a considerable amount of experimental data is obtained from these satellites, transmitting the data via satellite–ground links for subsequent analysis on the ground is often an inefficient process. Therefore, OEC can assist scientific satellites in completing on-orbit data processing, reducing unnecessary data transmission.

### *4.4. Satellite federated learning*

With the increasing demand for data-driven solutions in various fields, such as agriculture, healthcare, and environmental monitoring, federated learning has emerged as an attractive technique that addresses the challenges associated with centralized data processing and privacy concerns related to traditional machine learning methods. Satellite federated learning is a novel approach that combines the power of satellite technology and machine learning to enable efficient and privacy-preserving data analysis across geographically distributed data sources. For instance, in environmental monitoring, satellite federated learning can analyze multiple sources, such as satellite imagery, ground sensors, and weather stations, to monitor air and water quality, track deforestation, and predict natural disasters. By federating the data and applying machine learning algorithms, accurate models for environmental prediction and monitoring can be developed while protecting data privacy. Satellite federated learning expands more applications of OEC, such as multi-modality data fusion and efficient learning from across multiple satellites.[110] Salim et al.[111] proposed a comprehensive Federated Learning (FL)-based threat detection model for proactively recognizing intrusions in satellite communication.

We will now summarize three typical architectures of satellite federated learning, and discuss their characteristics. In Table 6,[44,73,112–119] we will also compare selected articles that may facilitate further study.

**Table 6** Comparison of satellite FL architectures.

| Objective | Name | Reference | Client | Server | Communication | Solution |
|---|---|---|---|---|---|---|
| LEO satellites generate a huge number of datasets, but they need to be transmitted to a GS, which is inefficient. | – | [120] | LEO | GS | no ISL | They firstly proposed satellite FL, and apply FedAvg [112] to show its excellent performance. |
| FL algorithms among satellites and GSs have challengs in unique trade-off between staleness and idleness. | FedSpace | [117] | LEO | GS | no ISL | They presented FedSpace, which considered a heuristic GS update procedure and gradient buffering to speed up convergence. |
| Based on the predictability of GS–satellite visiting lengths. The model staleness can be reduced by designing an appropriate scheduler. | – | [113] | LEO | GS | no ISL | They designed a scheduler based on the predictability of GS–satellite visiting lengths to reduce the model staleness. |
| No research talk about the intra-orbit ISLs in satellite FL. | – | [44] | LEO | GS | inter-orbit/ intra-orbit ISLs | They offered a comprehensive overview of the state of the art in satellite FL and outline several open research directions. |
| The costs of satellite communication can be reduced by FL, but there is no evaluation of these schemes. | – | [114] | LEO<br>GS | GS<br>LEO | intra-orbit ISLs | They proposed schemes considering the preliminary combination of FL and LEO satellite systems. The evaluation indicates that GS-LEO based method has better training performance than LEO-GS. |
| No research addresses the intrinsic properties of satellite constellations such as their non-visibility and predictability behavior. | – | [118] | LEO | GS/ MEO | intra-orbit ISLs | They leveraged the predictability of satellite movements and partial aggregating to greatly reduce the training time and communication costs. |
| Wireless channels between satellites and GS are highly unpredictable and unreliable. | FedHAP | [119] | LEO | HAP | intra-orbit ISLs, satellite-HAP, inter-HAP | They introduced high altitude platforms into a synchronous FL framework as server, and explore inter-satellite and inter-HAP collaborations. |
| Synchronous FL process can take several days to train a single FL model due to straggler satellites. | AsyncFLEO | [116] | LEO | HAP | intra-orbit ISLs, satellite-HAP, inter-HAP | They exploited the availability of synchronous FL without waiting for all satellites and tackle straggler satellites and model staleness. |
| The recognition of signal modulation in OEC and satellite FL can decrease communication costs and guarante data privacy. | – | [115] | UAV/ HAP/ LEO | MEO | UAV-MEO, HAP-MEO, LEO-MEO | They formulated a joint delay ratio and energy efficiency optimization problem and solved it by Q-learning. |

#### 4.4.1. *Satellite FL without ISL*

Razmi et al.[120] proposed using satellite FL based on LEO Mega-Constellation. This enables the data to stay in the satellite instead of being downloaded to the GS, saving time and energy. The Fl-based architecture typically consists of two planes: satellite and ground, and Chen et al.[114] explored their roles in FL, showing that GSs are more feasible for use as servers than LEO satellites. However, using LEO satellites as servers results in less delay and faster convergence. Jing et al.[115] introduce a new FL architecture that consists of UAV-LEO-MEO. Fig. 8(a) shows the architecture of the GS-LEO without ISLs, which is the basis of most studies.[116,117,120] The most significant drawbacks of this architecture are satellite idleness and model staleness because an update of the global model needs to wait for all of the client satellites to upload their local model.

#### 4.4.2. *Satellite FL with ISL*

Fig. 8 (b) illustrates the GS-LEO architecture with ISLs: client satellites can update the model directly if at least one of them has a connection to the GS.[44,118]

More research has been conducted on intra-orbit ISLs rather than inter-orbit ISLs because the latter are highly dynamic, making them vulnerable to Doppler shift and requiring more cautious planning of the connections.[121] Han et al.[122] aggregated models within device clusters and relayed models/-data to other satellites via ISLs, which ensures continuous FL despite limited satellite coverage time over specific areas.

LEO satellites have a short duration of visiting a GS, which is typically in the range of tens of minutes. Consequently, Jing et al.[115] replace GS with MEO to provide a longer visit time and reduce transmission delay and energy consumption. Nasrin et al.[123] placed the parameter server on LEO satellites at an altitude of 500 km to achieve more frequent interactions. Compared to GS as a parameter server, this method resulted in a 5× convergence speed and a 20% improvement in accuracy.

Additionally, they utilize HAPs and UAVs as an integral part of server side. This paradigm is illustrated in Fig. 8(c). Similarly, Cheng et al.[106] focus on a space/aerial-assisted computation offloading algorithm based on UAV-satellite-ground architecture.

#### 4.4.3. *GEO or aerial network as parameter server*

Figs. 8(a) and (b) are space/ground-assisted architectures, while Fig. 8(c) is a space/aerial-assisted architecture. The former is suitable for various earth observation applications, including urban planning, disaster management, and climate change mitigation.[124] It helps to avoid the transmission of large volumes of data to GSs. The latter has several advantages: it is less expensive, offers better visibility, allows better communications, and is easy to relocate.[119] Additionally, it can facilitate edge computing for UAVs and airplanes. Fig. 8 (c) shows the Air-LEO-GEO architecture, where the LEO-GEO configuration is useful for autonomous mega LEO satellite constellation management.[118] Tang et al.[125] introduced the potential advantages and applications of FL in SAGIN, discussing how FL can assist in solving resource scheduling problems in SAGIN. In summary, all these architectures aim to address the challenge of transmitting huge volumes of data from LEO satellite constellations, which leads to excessive spectrum occupancy.

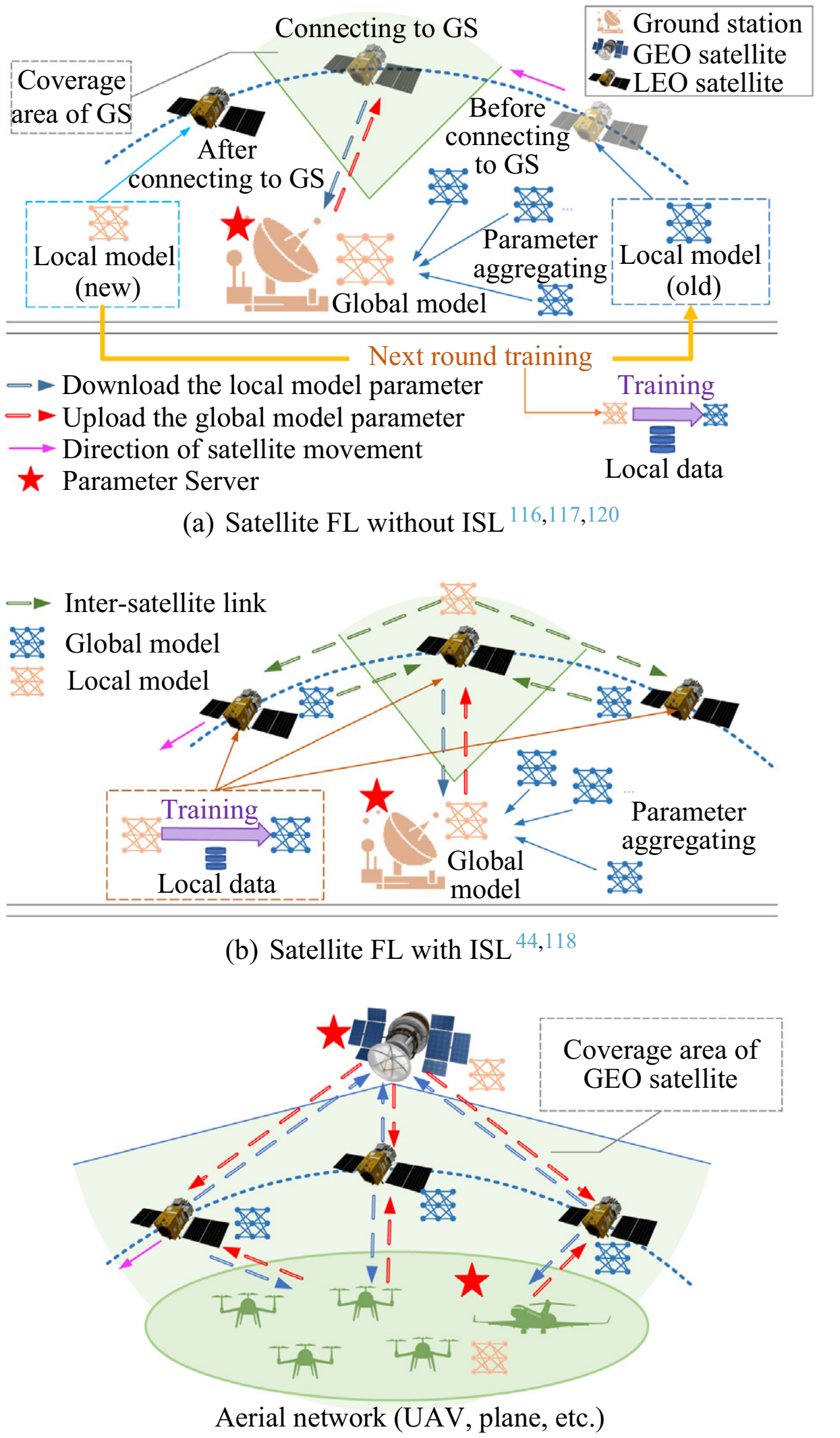

(a) Satellite FL without ISL[116,117,120]

(b) Satellite FL with ISL[44,118]

(c) GEO or aerial network as parameter server[106,118,124]

**Fig. 8** FL without and with ISLs in a satellite–ground system, and LEO-GEO system. In (a), satellites have to queue until they are within range of the GS, but in (b), satellites linked by ISLs can directly update the model as long as one of them is connected to the GS. In (c), GEO satellites perform a similar role as GSs, but offer superior coverage and benefit from the advantages of ISLs.[118]

#### 4.4.4. *Optimization in satellite FL*

The unique challenges of satellite FL include the dynamic nodes and communication bottlenecks, particularly in transmissions between different orbital planes and between satellites and the ground. Therefore, some research efforts focus on optimizing node selection and reducing communication pressure. Yang et al.[126] proposed two adaptive mechanisms for intra-plane and inter-plane model aggregation to accelerate the training process and integrated a self-compensation mechanism against transmission failure. Elmahallawy et al.[127] also proposed a new FL model aggregation scheme that optimally balances models between different orbits and orbit planes, and introduced a new communication topology that exploits HAPs to bridge satellites among different orbits to mitigate the Doppler shift.

Parameter transmission impacts the speed of federated learning, especially when the GS acts as the parameter server. Yang et al.[128] proposed a quantization-based method to accelerate parameter transmission, which highly accelerates the convergence by up to 2.8× and a 9.3× improvement in bandwidth utilization.

Client selection is another challenge in satellite FL because datasets are distributed across different satellite nodes, and data from some nodes may deteriorate the global training process. Wu et al.[129] proposed a client selection by investigating the motion characteristics of satellites in different orbits and introducing a metric to measure the value of each client in the overall training process. This method increased the model accuracy to 80%.

## 5. Simulators and testbeds

The emergence of cloud computing and edge computing has led to remarkable advances in multiple algorithms and applications. This progress has also brought about the development of exceptional simulation tools explicitly designed for cloud/edge computing, such as Edgecloudsim,[130] iFogSim,[131] and IOTSim.[132] However, as a nascent technology, OEC necessitates a more focused evaluation system and specialized simulation tools. This is because of the distinct differences between a satellite network and a ground network. Hence, this section compiles comprehensive details on the satellite configurations and simulation tools used in current OEC research. Our expectation is to provide guidance and to serve as a reference for future work in this field. Given the diverse configurations of satellite constellations, the same algorithm may introduce substantially different results when used in different scenarios. Furthermore, we have also summarized the evaluation systems and metrics utilized in relevant studies to provide guidance for the subsequent experimental design in this section. Finally, a testbed for OEC is introduced, which can provide a potential experimental platform for future work in this realm. Nevertheless, such research is still relatively limited in comparison with the abundant edge computing tools used for terrestrial applications. Hence, the testbed can be considered one of the numerous research opportunities in this field.

### *5.1. Satellite constellations for simulation*

The types of constellations used in satellite edge computing are still a topic worth discussing. However, in the face of the challenges of resource allocation and computation offloading in edge computing, some have actively explored possible constellations and conducted simulation experiments. Wang et al.[133] established a satellite-terrestrial double-edge computing system and simulated a satellite at an altitude of 500 km. Tang et al.[1] study the computation offloading problem between ground users and satellites based on three LEO satellites, each positioned at an altitude of 784 km. To investigate the effects of the number of satellites and of the altitude of the orbit on offloading and resource allocation in edge computing, Song et al.[134] simulated the performance of an algorithm for 4 to 6 satellites at altitudes ranging from 500 km to 1 500 km.

Iridium and Walker are common constellations applied in OEC. A constellation can be denoted by $i : h : t/p/f$, as previously explained in Section 3.1. The Iridium constellation, represented by $86.4^\circ : 780 : 66/6/1$, consists of 66 low earth orbit satellites evenly distributed in six orbits whose inclination is 86.4 degrees and height is 780 km. Based on the Iridium constellation, Wang et al.[135] and Jiang et al.[21] conducted research on computation offloading and resource allocation, respectively. Similarly, Wei et al.[20] focused on the utilization of edge computing in a satellite-based IoT with a $90^\circ : 1500 : 66/6/1$ constellation. Razmi[120] proposed federated learning by the satellites and conducted simulation experiments based on the $80^\circ : 500 : 25/5/1$ and $80^\circ : 2000 : 25/5/1$ constellations, and Elmahallawy used an $80^\circ : 2000 : 40/5/1$ constellation to study satellite federated learning in Refs. 116,119. Numerous studies have investigated the impact of the configuration of the constellation on the performance of an algorithm, including Tobias' research on the application of edge computing to improve the Quality of Service (QoS) for satellites. Pfandzelter et al.[136,137] performed simulation experiments based on four different constellation configurations: Starlink A ($53^\circ : 550 : 1584/72/1$), Starlink B ($81^\circ : 1275 : 375/5/1$), Kuiper A ($51.9^\circ : 630 : 1156/34/1$), and Kuiper B ($33^\circ : 630 : 784/28/1$). Most of the related work is based on LEO satellites, but there are also explorations of the possible use of MEO satellites. Kim et al.[138] analyzed OEC communication performance in the following three configurations: $48^\circ : 1400 : 63/7/1, 48^\circ : 800 : 190/10/1$ and $48^\circ : 20200 : 20/4/1$. However, the results indicate that MEO satellites are not a viable option for low latency services due to their long propagation delay.

In summary, LEO satellites are typically deployed at altitudes ranging from 500 km to 2 000 km for edge computing applications, with constellation sizes ranging from dozens to thousands of satellites, depending on the specific application scenario. However, increasing the number of satellites also increases the computational load on the simulations, which can quickly consume computing resources.

### *5.2. Experiment methods*

Just like edge computing on the ground, experiments on satellite edge computing primarily analyze network, communication, and computational performance. Current work focuses on optimizing resources, computation offloading, and exploring federated learning scenarios in satellite edge computing. This section outlines the usual experimental techniques and metrics employed in each of these domains.

#### *5.2.1. Resource optimization*

Resource optimization in satellite edge computing aims to minimize latency and energy consumption. Cui et al.[139] simulated a satellite equipped with a CPU capable of 1 000 cycles/bit, an energy consumption rate of $3 \times 10^{-10}$ J/cycle, and a communication capacity of 10 Mbps. Through their experiments, they explore the correlation between a satellite's computational capability and its task execution time. Moreover, Wang et al.[133] examines the commercial benefits of utilizing advanced resource allocation algorithms in OEC.

#### 5.2.2. Computation offloading

Computation offloading is an important strategy for improving the performance and energy efficiency of satellite communication systems. Metrics such as convergence speed, average latency, and energy consumption are often used to evaluate the effectiveness of offloading algorithms. For instance, previous work has explored the impact of task quantity and satellite computational ability on latency and has evaluated the performance of reinforcement learning algorithms in this context.[140,106,141] Other studies have focused on the energy consumption of ground users in scenarios where offloading involves interactions with them.[1] By considering factors such as the number of users, the computing and communication capabilities of the edge devices, and the types of tasks being offloaded, we can gain insights into the time and energy costs of different offloading strategies.

#### 5.2.3. Federated learning

As an efficient communication framework, FL avoids transmitting too much raw data, which mitigates the issue of limited communication bandwidth in the OEC of distributed computing.[120] Additional unique challenges being studied in this context are the short connection window and long connection cycle, which introduce problems of staleness and stragglers. This makes the speed of convergence and accuracy decrease.[117] Several efficient satellite FLs have been proposed and tested in public datasets such as MNIST and CIFAR.[120,116] A remote sensing image dataset FMoW is[142] widely used in satellite FL.[117] They compared the speed of convergence and the accuracy when the clients used Independent and Identically Distributed (IID) datasets and non-IID datasets. In our survey, an advanced method was able to achieve 87.79% accuracy (IID datasets) within 2:40 h and 82.9% accuracy (non-IID datasets) within 3:20 h by using CNN and MNIST.[116] Other research has aimed to accelerate an FL by conserving the communication bandwidth, particularly in the context of satellite–ground FL. Model quantization[143] is an approach that reduces the number of parameters during the interaction between the clients and the parameter servers.[114]

### 5.3. Simulation tools

In the field of edge computing, conducting experiments directly on physical infrastructure can be challenging and expensive, particularly for extensive systems like a satellite network. Thus, simulation tools such as virtual network simulators, edge computing simulators, and testbeds are crucial for evaluating the performance of edge computing in satellite networks. We summarize some of the available tools/testbeds that can be used for OEC simulation in Table 7.

#### 5.3.1. Satellite network simulators

The simulation of satellite networks plays a crucial role in OEC systems, as these networks' complex mobility patterns and distinct communication links are fundamentally unlike those in terrestrial edge computing. Currently, several mature tools for satellite network simulation exist, including SNS3,[144] STK, and MATLAB. SNS3 has been developed for satellite communication network simulation by leveraging the NS-3[4]. SNS3 is proficient in simulating various satellite link configurations and analyzing the network at the physical layer. Furthermore, Kassing et al.[145] proposed Hypatia, a network simulation tool, based on NS-3, specifically designed for simulating large-scale LEO constellations. It incorporates the distinguishing characteristics of LEO constellations, such as high-velocity orbital motion, to simulate and visualize their network behavior. Likewise, SILLEO-SCNS[146] concentrates on designing inter-satellite and satellite-GS networks and implements network visualization based on VTK. Lai et al.[147] introduced StarPerf, a comprehensive mega-constellation performance simulation platform. It empowers both constellation manufacturers and content providers to estimate and comprehend the performance achievable under various constellation options. Hence, it provides a tool for understanding and optimizing satellite networks.

It is exciting that in the research into SAGIN, these tools have demonstrated strong potential. Zhang et al.[148] proposed a platform to support SDN simulation in SAGIN, based on STK, ONOS and Mininet. Cheng et al.[77] presented a SAGIN simulation platform based on MATLAB, NS-3, STK and Python, which integrates multiple network protocols, node mobility, and control algorithms.

The successful implementation of these tools within SAGIN enables the analysis of the network and communication layers of OEC. Nevertheless, there is an evident lack of a comprehensive OEC simulator or testbed built upon these existing tools. It has become clear that to bring our OEC simulator to a level closely aligned with actual satellite networks, we require such professional network simulation tools.

#### 5.3.2. Testbeds

At the time of writing, Celestial[19] is the only testbed available for satellite edge computing. It uses the SILLEO-SCNS network simulator to simulate large-scale LEO constellations while simulating edge computing based on microVMs. One unique aspect of the testbed is its ability to execute a high-definition video conference task based on the Starlink satellite constellation, with servers deployed across South Africa, Cameroon, Nigeria, and Ghana. The results show that OEC effectively reduces the latency in the meeting. Celestial also provides support for simulation related to OEC challenges, such as resource allocation, computation offloading, and satellite routing.

Celestial is a successful example: it provides a feasible application simulation under the OEC scenario. Nevertheless, simulations closer to the real hardware for OEC will provide more opportunities for OEC, such as integrating with real computing hardware and communication equipment, which is beneficial to those who are committed to OEC energy management and communication link optimization. At the same time, we hope to consider the harsh environment of satellites in space, which we have discussed in Section 3.1. The simulation of the effects of radiation on the OEC hardware poses a significant challenge. However, this is important when considering the practical applications of an OEC system, as such issues cannot be overlooked.

[4] https://www.nsnam.org/

Table 7 Available simulation tools/testbeds for satellites and OEC.

| Tools/ Testbeds | Characteristics/Focus | Links |
|---|---|---|
| MATLAB | Satellite Communications Toolbox: Configuration, Simulation, Measurement, and Analysis of end-to-end satellite communications links. | ww2.mathworks.cn/en/products/satellite-communications.html |
| STK | Realistic and time-dynamic three-dimensional simulation: satellite orbit perturbation, satellite attitude, coverage, access, network topology, satellite communication links, etc. | www.ansys.com/products/missions/ansys-stk |
| SNS3 | Physical level burst resolution dynamic system simulation: visualization and analytics for the simulation results complemented by scalables cloud computing capacity for running simulations. | https://satellite-ns3.com/ |
| StarPerf[147] | A mega-constellation performance simulation platform: area-to-area network performance profiling, support for scaling the constellation and space resource (e.g., number of satellite, link availability and capacity) | https://github.com/SpaceNetLab/StarPerf_Simulator |
| Celestial[19] | A simulator support for OEC based on microVM: accurate emulation of LEO constellations, support for arbitrary software systems and cost-efficient scalability. | https://github.com/OpenFogStack/celestial |
| ONOS | An open-source SDN platform. It supports features like traffic engineering, virtualization, and network slicing. | https://opennetworking.org/onos/ |
| Mininet | A network emulator to create virtual networks. It can simulate network topologies, test network protocols, and develop SDN applications. | http://mininet.org/ |
| microVMs | An open-source virtualization technology that creates lightweight virtual machines, providing a secure and efficient way to build, deploy, and manage multi-tenant container and function-based services. | https://firecracker-microvm.github.io/ |

## 6. Algorithms

Computation offloading, the process of transferring computational tasks from resource-constrained devices to more powerful servers,[149] is crucial in optimizing resource utilization and enhancing the performance of OEC systems.[1] Effective offloading strategies enable seamless integration between terrestrial IoT devices and satellite-based resources, enabling low latency, high throughput, and reliable task execution.

The computation offloading of OEC can be described as $M$ OEC servers and $N$ clients, which can be denoted by $\mathscr{M} = \{1, 2, \cdots, M\}$ and $\mathscr{N} = \{1, 2, \cdots, N\}$, respectively. Assume that each OEC server $m$ is equipped with computational capability $f_m^S$, which denotes the clock frequency of the CPU. Each client $n$ has the computation task $W_n \triangleq (D_n, X_n)$, where $D_n$ is the amount of the input data and $X_n$ is the number of CPU cycles required by the task $W_n$. The client $n$ can execute tasks locally or offload them to an OEC server or terrestrial cloud server. Let $a_{n,m} \in \{0, 1\}$ indicate whether $W_n$ is offloaded to OEC server $m$, and $b_{n,m} \in \{0, 1\}$ indicate whether it is offloaded to terrestrial cloud server.[1]

The majority of currently available offloading algorithms focus on minimizing latency and energy consumption. This latency includes the propagation delay, transmission delay, and computing delay. Some studies may also take into account the queuing waiting time for tasks. Energy consumption can be characterized as the aggregate energy cost of the client devices, OEC servers, and terrestrial cloud servers.

Then, the latency in a round of computation offloading can be described as follows:

$$\begin{aligned} T_{n,m} &= (1 - a_{n,m} - b_{n,m}) \tfrac{X_i}{f_n^L} \\ &+ a_{n,m} \left( \tfrac{s_{n,m}}{c} + \tfrac{D_i}{R_{n,m}} + \tfrac{X_i}{f_{n,m}^S} \right) \\ &+ b_{n,m} \left( \tfrac{s_{n,m} + s_{g,m}}{c} + \tfrac{D_i}{R_{n,m}} + \tfrac{D_i}{R_{g,m}} + \tfrac{X_i}{f_{n,g}^G} \right) \forall n, m \end{aligned} \tag{5}$$

where $s_{n,m}$ is the distance between client $n$ and OEC server $m$, and $s_{g,m}$ is the distance between OEC server $m$ and terrestrial cloud server $g$. The constant $c = 3.0 \times 10^8$ m/s is the speed of light, while $R_{n,m}$ is the channel capacity between client $n$ and server $m$ and $R_{g,m}$ is that between client $n$ and terrestrial cloud server $g$. The clock frequency of the CPU of the server $m$ allocated to client $n$ is $f_{n,m}^S$, while that of the terrestrial cloud server $g$ is $f_{n,g}^G$.

The energy consumption of a round of computation offloading is

$$\begin{aligned} E_{n,m} &= (1 - a_{n,m} - b_{n,m}) E_n^L + a_{n,m} (E_{n,m}^{ST} + E_{n,m}^{SC}) \\ &+ b_{n,m} (E_{n,m}^{GT} + E_{n,m}^{GC}) \qquad \forall n, m \end{aligned} \tag{6}$$

where $E_n^L$ is the energy consumption of client device $n$ when task $W_n$ is executed locally, $E_{n,m}^{ST}$ and $E_{n,m}^{SC}$ are respectively for the transmission and computation energy consumption of OEC server $m$ when task $W_n$ is offloaded to server $m$. $E_{n,m}^{GT}$ and $E_{n,m}^{GC}$ are respectively the transmission and computation energy consumption of terrestrial cloud server $g$ when task $W_n$ is offloaded to terrestrial cloud server $g$.

Compared to terrestrial MEC, OEC provides an important supplement for terrestrial IoT. However, the characteristics of the satellites introduce unique challenges for computation offloading and resource management, such as the intermittency of communication. In our survey, the computational services provided by satellites can serve ground users, HAP, and

**Table 8** Comparision of algorithms - convex optimization.

| Ref. | Offloading decision | Optimization objective (latency) | Optimization objective (energy) | Methods | Computational complexity | Performance (cost reduction) |
|---|---|---|---|---|---|---|
| 150 | Satellites to ground servers or other satellites | propagation (U-S, S-S), transmission (U-S, S-S) and computation (G,S) | transmission (U-S,S-S) and computation (S) | Double edge computation offloading algorithm | (not provided) | Latency: 44.88% Energy: 49.22% (compared to fixed polling mode.) |
| 1 | Ground users local computing or offloading to satellites or ground servers | propagation (U-S, S-S), transmission (U-S, S-S) and computation (U,S,G) | computation (U,S,G) | Binary variables relaxation method and alternating direction method of multipliers | $O(I^3) + O(I(M+1)) + O(I)$ | Latency: 16.5% Energy: 15.7% (compared to offloading to cloud servers.) |
| 134 | Ground users local computing or offloading to satellites | propagation (U-S), transmission (U-S) and computation (S) | transmission (U-S) and computation (U,S) | Lagrangian dual decomposition method and sequential fractional programming based Dinkelbach's algorithm | polynomial | Energy: $\sim$ 60% (compared to local computing) |
| 153 | Ground users local computing or offloading to satellites | propagation (U-S), transmission (U-S) and computation (S) | (not considered) | Grey Wolf Optimizer and Lagrange multiplier method | $O(N^W NI)$ | Latency: $\sim$ 30% (compared to random offloading decision) |
| 154 | Satellites to other satellites | transmission (S-S) and computation (S) | transmission (S-S) and computation (S) | Successive convex approximation method | (not provided) | Energy: $\sim$ 42% (compared to equal resource allocation) |
| 93 | Ground users local computing or offloading to satellites | waiting (U), propagation (U-S), multi-hops transmission (U-S) and computation (S) | transmission (U-S) and computation (S) | Hybrid genetic binary particle swarm optimization algorithm | $IU \cdot O(N)$ | System Cost: $\sim$ 13.2% (compared to genetic algorithm) |
| 151 | Ground users offloading to HAP and HAP offloading to satellites | transmission (U-HAP, HAP-S) | computation (U,HAP,S) | Quadratic transform based fractional programming, quadratic transform based fractional programming and interior point method | $O(L_0 L_1 I^3 K^3 + L_0 L_2 (I\vert\mathscr{K}\vert Nd + \vert\mathscr{K} Mb\vert)^3 + L_0 L_3 I^3)$ | Energy: $\sim$ 51.7% (compared to MU-MIMO computation offloading scheme[152]) |

Ref. is the reference's number (the same below).
"U-S" and "S-S" respectively represent the links between ground users and satellites, and the inter-satellite links. "G", "S,","U", and "HAP" respectively represent the ground servers, satellite edge servers user devices, and aerial high altitude platforms (the same below).
$I$ denotes the maximum number of iterations, $N$ is the number of users (IoT devices or satellites), and $M$ is the number of servers (the same below).
$N^W$ in Ref. 153 is the number of grey wolves in the population.
$U$ is the population size in the ABPSO algorithm. [93]
The indices of latency and energy cost in performance (column 7) are all average values.

**Table 9** Comparision of algorithms - game theory.

| Ref. | Offloading decision | Optimization objective (latency) | Optimization objective (energy) | Methods | Computational complexity | Performance (cost reduction) |
|---|---|---|---|---|---|---|
| 135 | Ground users local computing or offloading to satellites | transmission (U-S) | computation (U,S) | Classical bisection method and Lagrange multiplier | $O(KMN(log(I/\varepsilon)))$ | System Cost: ∼ 75% (compared to random offloading decision) |
| 162 | Ground users local computing or offloading to satellites | transmission (U-S), propagation(U-S), computing (G,S) | transmission (U-S), computation (U,S) | Gale-Shapley algorithm, coalition game iterative algorithm, Rosen gradient projection method and Lagrange multiplier method. | $2NM(\frac{NI}{\varepsilon_1}+NMK)$ | System Cost: ∼ 26% (compared to random offloading decision) |
| 163 | Ground users local computing or offloading to UAVs, satellites and ground servers | transmission (U-S, U-UAV,U-G) propagation (G-S,G-UAV), computing (U,G,UAV,S) | transmission (U-G, U-UAV,U-S), computation (U,G,UAV,S) | Multi-leader and multifollower Stackelberg game and exact potential game | (not provided) | Latency: 24% Energy: 32% (compared to random offloading decision) |

$K$ is the number of iterations and $\varepsilon$ the accuracy requirement in Ref. 135.
$\varepsilon_1$ indicates the Rosen gradient projection method and $K$ the maximum number of iterations of the Lagrange multiplier method in Ref. 162.
In Ref. 151, $K$ is the number of HAPs, and each HAP is equipped with $M$ satellites, $I$ is the number of users and each user is equipped with $N$ antennas, $d$ and $b$ are the data streams supported by each user and HAPs, respectively. $L_0, L_1, L_2$ is the number of iterations for the three satellites' on-orbit tasks. Their common issue lies in the limited resources for satellite services and highly dynamic communication links, aiming to ensure the success rate of tasks, reduce service delay, and minimize energy consumption.

**Table 10** Comparision of algorithms - heuristic algorithm.

| Ref. | Offloading decision | Optimization objective (latency) | Optimization objective (energy) | Methods | Number of iterations to convergence | Latency performance compare to random strategy |
|---|---|---|---|---|---|---|
| 155 | Ground users offloading to GS or satellites | propagation (U-S, S-S), transmission (U-S, S-S) and computation (G,S) | transmission (U-S,S-S) and computation (S) | K-medoids clustering and genetic algorithm | ∼ 50 | 53.79% |
| 156 | Ground users local computing or offloading to satellites | transmission (U-S, S-S) and computation (G,S) | computation (U, S) | A low-complexity heuristic algorithm | ∼ 20 | (not provided) |
| 157 | Ground users offloading to satellites | propagation (U-S, S-S), transmission (U-S, S-S) and computation (G,S) | (not considered) | Genetic algorithm | (not provided) | 43.34% |
| 158 | Ground users offloading to UAVs or satellites | transmission (U-S/UAV, S-S/UAV) and computation (S,UAV) | (not considered) | Particle swarm optimization and genetic algorithm | ∼ 500 | 44.70% |

In this section, we survey the algorithms for computation offloading and resource management. They are divided into four categories: convex optimization, game theory, heuristic algorithm, and deep learning. We also compare the scenarios, methods and performance of these algorithms in Table 8,[1,93,150–154] Table 9, Table 10[155–158] and Table 11[94,139,141,159–161] and discuss their advantages and disadvantages in Table 12.

### 6.1. Convex optimization

Convex optimization provides a well-established framework for solving optimization problems with convex objective functions and constraint sets. These methods can be effectively applied to address the computation offloading problem in OEC systems, where the goal is to minimize the overall latency or energy consumption while satisfying task deadlines and satellite edge servers' resource constraints. Hence, the objective function can be formulated as the overall latency,[153] the energy consumption,[1,134,154] or a linear combination of them.[93]

The offloading decision can be formulated as a mixed-integer linear programming problem, where integer variables represent the mapping of tasks to computing nodes. This is harder to solve than the continuous problem. Tang et al.[1] fixed the problem by using the binary variables relaxation method and proposed a distributed algorithm by leveraging the alternating direction method of multipliers. The results have shown that the method can reduce the total energy consumption of ground users by 16.3% and 31.8% compared with fully offloading and local computational, respectively.

**Table 11** Deep learning.

| Ref. | Offloading decision | Optimization objective (latency) | Optimization objective (energy) | Methods | Number of iterations to convergence |
|---|---|---|---|---|---|
| [141] | Ground users to satellites or ground servers | transmission (U-S) and computing (G,S) | transmission (U-S) and computing (G, S) | Deep Reinforcement learning (DRL) | ~ 10 000 |
| [139] | Ground users to satellites or ground servers | waiting (G,S), transmission (U-S), propagation(U-S) and computing (G,S) | computing (S) | Lagrange multiplier method and DRL | ~ 10 000 |
| [159] | Ground users local computing or offloading rightarrow satellites | transmission (U-S) and computing (G,S) | computing (G,S) | DRL with multiple parallel deep neural networks (DNN) | ~ 300 |
| [94] | Ground users local computing or offloading rightarrow satellites, ground servers and waiting for next scheduling | computing (U,G,S) | computing (U,G,S) | DRL with deep deterministic policy gradient (DDPG) algorithm | ~ 2 500 |
| [161] | Ground users local computing or offloading to satellites | transmission (U-S), propagation(U-S) and computing (G,S) | computing (U,S) | DRL with attention mechanism and proximal policy optimization collaborative algorithm | ~ 25 |
| [160] | Ground users local computing or offloading to LEO/MEO/GEO satellites | transmission (U-S), propagation(U-S) and computing (G,S) | transmission (U-S) and computing (G,S) | Dynamic distributed learning offloading algorithm based on distributed deep learning. | ~ 300 |

**Table 12** Advantages and disadvantages of algorithms.

| Algorithm | Advantages | Disadvantages | When to use |
|---|---|---|---|
| Convex optimization | Theoretically guarantees finding the optimal solution. | Requires accurate modeling of computation tasks and OEC servers. | The models of computation tasks and MEC servers can be accurately built and the problem is small-scale. |
| | Relatively simple algorithm implementation. | Solving time can be long for large-scale problems | |
| Game theory | Considering the conflict of interests between users and OEC servers | | The conflict of interests between clients and OEC servers is significant and the problem is large-scale. |
| | Finding equilibrium strategies for each player | Algorithms for solving game problems are often complex. | |
| Heuristic algorithm | Designing with low-complexity and cost-effective to implement. | Typically find local optima and cannot guarantee global optimality. | The problem has a large search space, an approximate solution is acceptable, and computational resources are limited. |
| | Providing good approximate solutions when dealing with problems with very large search spaces. | Parameter tuning demands intensive experimentation. | |
| Deep learning | Ability to learn complex constraints and relationships between clients and servers. | Requires a large amount of data for training | The relationship between clients and OEC servers is complex and data is abundant. |
| | Ability to handle non-linear and discrete problems. | The training process can be time-consuming and hard to find the optimal solution. | |

Computation offloading in OEC has complex constraints, such as the resource constraints, the deadlines of the tasks, and which satellites are usable, which makes the problem intractable. A feasible solution is to decompose the problem into several sub-problems. Thus, Song et al.[134] applied the Lagrangian dual decomposition method to solve the subproblems. Wang et al.[154] take the first-order Taylor expansion to replace concave items and propose a successive convex approximation method to deal with them. The results show that the method of Ref. 154 can reduce the total energy consumption by 28% and 68% compared with collaborative equal resource allocation and non-collaborative optimized resource allocation, respectively.

Furthermore, Hu et al.[93] considered inter-satellite communication and multi-hop routing in computation offloading and proposed a hybrid genetic binary particle swarm optimization algorithm to solve this. The proposed method increases the convergence speed by 48% compared to the genetic algorithm and reduces the total system consumption (composed of energy and transmission) by 10%.

### 6.2. Game theory

Game theory provides a powerful framework for analyzing and designing interactions between rational decision-makers, such as IoT devices and satellite edge nodes in OEC systems. In this context, game theory can be used to model the offloading decision as a game, where each player (i.e., client device) will selfishly choose the strategy that will get a better QoS by deciding whether and where to offload a task.

Wang et al.[135] proposed a game-theoretic framework for computation offloading in OEC. Within this framework, the response time and energy consumption of a task, calculated using queuing theory, serve as metrics for optimization. The existence and uniqueness of a Nash equilibrium for this game are established through a rigorous theoretical analysis. Furthermore, an iterative algorithm is presented to efficiently compute the Nash equilibrium. The results indicate that the offloading decision under Nash equilibrium yielded an average cost saving of 82% when compared with local computing, 50% compared with even offloading, and 82% compared with random offloading.

If each satellite can match at most number of clients and each client can match at most one satellite, the problem can be modeled as a many-to-one matching model, and Wang et al.[162] solved it by the improved Gale–Shapley algorithm and a coalition game iterative algorithm. Their experiments showed that the system cost of matching-coalition offloading decision is reduced by 15.32%, 22.89%, and 37.89% on average compared with greedy offloading, random offloading, and total offloading decisions, respectively.

Lin et al.[163] proposed a multi-leader and multi-follower Stackelberg game, which includes a leader subgame for the OEC servers and a follower subgame for the edge nodes. They proved that a Stackelberg equilibrium exists in the proposed method. In this game, the OEC servers are the leaders, and the client nodes are the followers. The leaders set the charge prices for computational resources, and the client nodes should select appropriate offloading links and task offloading ratio. Their methods reduce by approximately 36%, 50%, and 60% the average energy consumption compared with random offloading, fully offloading, and local computing, respectively.

### 6.3. Heuristic algorithm

Heuristic algorithms are vital in addressing complex optimization problems where traditional methods might fall short due to the size and complexity of the search space. In the context of OEC, heuristic algorithms can be particularly useful.[78] The dynamic and resource-constrained environment of satellites requires efficient decision-making processes for tasks such as data processing, resource allocation, and communication scheduling. Heuristic methods, such as genetic algorithms,[157] simulated annealing [164], or particle swarm o optimization,[158] can provide adaptive and robust solutions by exploring a variety of potential solutions and iteratively improving them. This adaptability is crucial in OEC scenarios where conditions can rapidly change, and quick, effective responses are essential to maintaining system performance and reliability.

The first step of heuristic algorithms is defining appropriate objectives and constraints, followed by generating an initial set of solutions, which can be either randomly generated or based on specific heuristic rules. Subsequently, the objective function value of each solution is calculated to evaluate the quality of the solutions. The critical designing of these algorithms is the search strategy and the acceptance criterion, and the process iterates until a termination condition is met.[165] For example, in the simulated annealing algorithm, the search strategy involves randomly selecting a new solution in the neighborhood of the current solution. If the new solution is better than the current one, it is accepted unconditionally; otherwise, it is accepted with a probability of $e^{\Delta E/T}$, where $\Delta E$ is the difference in the objective function value, and $T$ is the current temperature.

In OEC, the inputs of heuristic algorithms can include the set of all satellites, the satellite network topology, and the set of tasks. The outputs can be the offloading strategy, task execution time, and the number of overtime tasks.[157] For example, in the genetic algorithm, the population (the set of each task offloading strategy) is initialized randomly, and the fitness value of the population is calculated to evaluate the results of the current iteration. The fitness is the optimization objective, which can be task delay, energy consumption, or a combination of both. In each iteration, the following steps are executed: (A) generating a new population by selecting the better individuals (task offloading strategies); (B) performing crossover operations with probability $p_c$ (exchanging part of the tasks' allocation between two strategies); (C) performing mutation operations with probability $p_m$ (randomly changing the allocation of tasks). The iteration continues until the maximum number of iterations is reached or the fitness requirement is met.[166]

The heuristic algorithm is an efficient approach for addressing the computational offloading optimization problem in OEC. They can obtain feasible solutions in a short time without much computational resources. However, the efficiency of heuristic algorithms depends on the quality of the initial solutions and the effectiveness of the search strategy, which makes them lack universality across different OEC optimization problems.

### 6.4. Deep learning

Deep learning has emerged as a powerful tool for solving complex optimization problems, including the computation offloading problem in OEC systems. Deep learning is very good at facing some complex optimization functions. By designing reasonable loss functions, the optimal solution can be gradually approached in an iterative process.

The framework of Deep Reinforcement Learning (DRL) consists of agents, policies, actions, the environment, and a reward function. In each time slot, agents (clients) take actions (offloading decisions and resource allocation) according to the current state (ongoing tasks, unsevered tasks, communication resources, computing resources, etc.) of the system. The reward function is designed to express the long-term reward by evaluating the difference between the current environment and the ideal environment so the method can then update the policy to make better decisions for the agent in the next training round.[141] Cui et al.[139] proposed a DRL algorithm to obtain the optimal computing and communication resource allocation for every time slot. They model the joint user association and offloading decision problem as a Markov Decision Process (MDP) with large state and action spaces. Their experimental results show that it can reduce the total latency by approximately 13% compared with random offloading and 23% compared to greedy computing.

Deep Deterministic Policy Gradient (DDPG) is an algorithm consisting of several different deep neural networks. This method addresses the problem of simultaneous learning of discrete variables (offloading decision) and continuous variables (resource allocation). The DDPG has an actor-network design and a critic network, whose functions are executing decisions and guiding the correctness of the behavior, respectively. Zhang et al.[94] applied this method in a proposed three-tier OEC architecture, and extensive simulation results have shown that DDPG improves the reward by nearly three times and average energy consumption by nearly 50% compared with a deep Q-Network-based algorithm.[167]

The complex environment and numerous variables affect the convergence and performance of DRL. Therefore, introducing an attention mechanism can let the agent focus on the most important variables in the environment, thus improving its convergence. At the same time, in the highly dynamic network topology of OEC, the DRL, with the attention mechanism introduced, can adapt to environmental changes more quickly. Chai et al.[161] considered the interdependencies between the tasks and their different priorities and modeled them as Directed Acyclic Graphs. They introduced an attention mechanism in the DRL and proposed the proximal policy optimization collaborative algorithm to learn the best offloading strategy. Their results showed that their proposed method can reduce costs by 8.87%, 14.30%, and 18.18% compared with greedy offloading, HEFT algorithm, and even offloading, respectively. (HEFT algorithm first ranks the tasks based on priority and then schedules each task with the lowest offloading cost.).

## 7. Issues, challenges, and future research directions

For OEC, the uniqueness of the environment in space brings many new challenges. In this section, we consider the following research challenges which will need to be addressed, and enumerate opportunities for further academic research.

### 7.1. Issues and challenges

The vision of a reliable and accessible orbital edge computing service faces numerous challenges.

#### 7.1.1. Limited orbital resources

The main source of energy for the satellites comes from solar power. During flight, they cannot obtain energy half of the time, due to the earth's shadow, when they will have to rely on their own energy storage. However, with the miniaturization of satellites, there are physical constraints, such as limited weight and volume.[17]

#### 7.1.2. Harsh environment in space

Pfandzelter et al.[81] present a taxonomy of failures that may occur in OEC and explain their effects on the software system in a satellite. As we mentioned in Section 3.1, the harsh environment of space poses a huge challenge for the software and hardware design of satellites in OEC.[85,168]

#### 7.1.3. Launch cost limitations of satellite hardware

High-performance computing hardware is often more vulnerable to radiation, while also generating higher energy consumption and thermal power dissipation. In space, heat can only be dissipated by radiation after conduction to the surface, as convection is not possible. Therefore, adding radiation shielding, heat dissipation equipment, and more energy storage to satellites significantly increases their size and weight, making launch costs expensive. For instance, SpaceX's Falcon 9 launch costs around $62 million to deliver a 22 800 kg LEO satellite, $2 720/kg.[169]

#### 7.1.4. Increased density of satellites in orbit

Recently, large-scale LEO satellite constellations have become a prominent trend in intelligent satellite networks. The massive scale of a satellite network presents challenges to optimization algorithms. There are now more than 3 558 Starlink satellites in orbit. The satellites perform different computing and storage tasks, each with different levels of resource consumption. Hence, optimizing the computing and communication resources of satellite clusters in a large constellation is a challenge.[170]

#### 7.1.5. Managing high mobility

The high mobility of LEO satellites causes the connections between satellites and the ground network to constantly change. This frequent switching of satellite–ground links brings challenges to mobility management in OEC services. At the same time, ground users are not limited to GSs: personal terminals, aircraft, and ships also have a certain mobility. Mobility at the user end brings more challenges to mobility management.[34]

#### 7.1.6. Non-ideal simulators

At present, research on OEC is restricted to ground simulation, which might not represent the actual environment in space. Satellite constellations are still under construction, and we also lack real public constellation telemetry data for OEC simulation.[19] The mission of the "TianSuan" series satellites[9] is to create a computing platform in space, providing an opportunity for OEC researchers. Nevertheless, performing experiments on actual satellite equipment continues to be expensive. Moreover, achieving OEC under the envisioned "large-scale constellation" seems like a distant prospect.

#### 7.1.7. Unclear mechanisms and structures

OEC is still in the concept stage, with only a few studies on real satellites.[9] However, there is still a lack of design for the OEC architecture of a fully functional constellation. There has been little consideration of the distribution of tasks and collaboration between different types of satellites, such as communication, remote sensing, and navigation satellites. In addition, there is a lack of systematic description and research on the design of a constellation and the network topology for OEC.

#### 7.1.8. Security

The openness of the electromagnetic environment in OEC brings security problems. The cybersecurity and space activities are intricately connected. Satellite communication channels are vulnerable to interception, eavesdropping, and jamming. This requires an OEC system to have the capability of threat detection, defense, and data privacy preservation.[111] Fortunately, emerging technologies such as federated learning,[111] semantic communication,[102] and quantum cryptography offer promising solutions to OEC security challenges.

### 7.2. Opportunities and future research

There are high hopes for the applications of OEC, including a trend to make satellite networks more intelligent. There are numerous opportunities for researchers who seek to contribute to its continued development. We will now summarize the current work and proposed potential research opportunities.

#### 7.2.1. LEO satellite network design

The LEO satellite constellation is critical to the future development of 6G networks.[171] However, the design of a satellite constellation for OEC services has not yet been discussed. The rapidly developing commercial LEO satellite constellations, such as StarLink, mainly undertake communication services and have not claimed to provide OEC services. The design of an OEC constellation needs to consider factors such as coverage area, communication capabilities, and inter-satellite networking, but at present, there is no specific work that describes the constellation design under OEC.

#### 7.2.2. Resource optimization

Limited bandwidth and on-board processing capacity significantly impact the performance of OEC. Therefore, selecting an appropriate routing strategy and node selection to maximize the utilization of these resources is a key challenge.[172] This is a complex optimization task that requires considering the time-varying resource status of each satellite, the spatial changes of the entire topology network, and link and node failures in order to enhance the fault tolerance capability of the OEC system.

#### 7.2.3. Space data center

Migrating storage resources closer to the user end can greatly reduce latency and energy consumption. Satellites generate massive amounts of data through daily earth observation. Therefore, the construction of a satellite data center can avoid a large amount of data transmission and thus save communication resources. Furthermore, a space data center is less vulnerable to earthquakes, hurricanes, and tsunamis. Peng et al.[173] proposed an architecture that integrates the data center into the distributed satellite cluster network and explains the benefit to OEC.

#### 7.2.4. Satellite caching strategies

With SDN and NFV in place, communication and computing functionalities are converging in terrestrial communication networks.[174] Using satellite caching technology to improve the efficiency of the distribution of content and reduce business latency is a promising method.[175,176] However, due to the dynamic topology of a satellite network, traditional high-speed caching, and allocation schemes are not suitable, because the coverage areas of different satellites are dynamically changing in real time, which increases the difficulty of designing caching schemes. Hence, cooperative caching of satellites is a challenging research area.[177] Leveraging edge computing to enhance content caching can become an attractive research direction for OEC, and the computing power deployed at the satellite edge can make content services more efficient and flexible.[178]

#### 7.2.5. Computation hardware and server design for OEC

Computational hardware is a key issue worth attention in OEC, and the surging OEC business demand is stimulating the development of more advanced space computing hardware. With the entry of COTS servers into space, designing OEC hardware with high energy efficiency and reliability is crucial.[81] Designing OEC servers is extraordinarily different from the edge computing servers on the ground. At the rocket launching time, the killing factors of the server are its weight and volume. Due to the harsh environment in space, hardware hardening and redundancy techniques have been widely adopted. Xu et al.[86] have proposed a novel satellite-born server architecture, namely SoC–Cluster, that consists of massive, low-power SoCs. It outperforms conventional servers such as Intel CPU and NVIDIA GPU.

#### 7.2.6. Virtualization

Virtualization can provide a way to address the challenges related to service deployment on OEC. Guo et al.[34] have proposed lightweight virtualization and application in the OEC, including lightweight application architectures such as microservices that can enable the rapid deployment of applications.[179] They emphasized that the limited resources of satellites require further research to adapt containers and go serverless or develop a new lightweight virtualization technology, and corresponding lightweight virtualization resource management tools, like KubeEdge and Kubernetes. Virtualization is also important for OEC simulators, such as Celestial,[19] an OEC testbed based on a lightweight virtual machine MicroVM. We need a virtualization tool for OEC that can be applied to real satellites, and can also be applied to the OEC testbed to carry out various experiments with algorithms, such as computation offloading, resource management, etc.

#### *7.2.7. Containerization*

Containerization is a promising technology that can be applied to OEC. Containerization packages a software application and its dependencies into a lightweight, portable container. The computational payloads on satellites can include FPGAs, CPUs, GPUs, etc. Containerization can unify these hardware resources, increasing the flexibility of OEC services in complex scenarios.[180]

#### *7.2.8. Microservice*

Microservices are a software development approach that structures an application as a collection of small, independent services.[181] The self-contained, lightweight, and portable attributes of microservicemakees them a perfect fit for satellite networks characterized by limited resources and rapidly changing topologies. The architecture of microservices can promote task migration and computation offloading on satellites, along with GS detection and satellite management in OEC.[182] Wang et al.[58] have applied cloud-native technology in the LEO satellite “BUPT-1”, which is based on microservice architecture. The deployment of microservices in an OEC platform can provide more flexible and efficient management, and enhance the scalability and reliability of OEC.

#### *7.2.9. Resilient network slicing*

Resilient network slicing is a promising virtualization technique where the infrastructure is shared by multiple tenants (operators) to serve multiple service classes simultaneous.[183] Esmat et al.[184] identified possible types of failures in satellite-terrestrial edge computing networks, and present solutions to enhance the resilience of communications and computing.

#### *7.2.10. Federated learning*

As an important method for computational privacy protection, federated learning is widely used in terrestrial MEC.[185] In recent years, satellite federated learning has also been studied with the development of OEC;[120,117] it mitigates the limited resources of a single satellite while protecting data privacy. The challenge of satellite federated learning is that each round of training may accept results from different clients, and the quality of the data is unpredictable, which could worsen the training process.[129] Current discussions on satellite federated learning mostly focus on designing federated learning mechanisms based on the mobility of satellites, ignoring the issue of satellite data quality distribution in actual application scenarios.

#### *7.2.11. Testbed*

At present, OEC is still a concept, and only one comprehensive testbed work on OEC has been proposed.[19] However, due to the large number of nodes in satellite constellations and the complex network topology, developing OEC models on real devices is not feasible. Hence, with the diversification of application scenarios and the introduction of new algorithms, there is an urgent need to develop more testbeds and improve their performance. A test-friendly OEC testbed model should have the following characteristics:

(1) Accurate satellite constellation simulation. Satellite network topology and satellite mobility models are one of the most important features of OEC tasks. Accurate satellite constellation simulation can realistically simulate the movement of the nodes in space. Typically, it can support professional satellite simulation tools such as STK, or develop independent satellite constellation simulation tools.
(2) Excellent platform compatibility. The testbed should be deployable on actual physical devices to achieve a more realistic simulation. Therefore, the testbed needs to be compatible with various platforms, such as personal computers and various embedded devices.
(3) Customizable resource management. Limited satellite resources have become a recognized challenge. The testbed should support the simulation of satellite edge computing performance under various resource conditions.
(4) Support for actual task execution. OEC is designed for a variety of services, such as deep learning training and inference, channel encoding and decoding, intelligent task scheduling, etc. The testbed should support simulations of these actual applications.

## 8. Conclusions

This paper has introduced the definition and architecture of Orbital Edge Computing (OEC), explored the differences between OEC and Mobile Edge Computing (MEC), described the challenges of implementing edge computing on satellites, and outlined potential technologies that can be utilized. Additionally, this article has examined the unique characteristics of satellite networks and their application in OEC, summarized key technologies such as network design, computation offloading, and resource allocation, and identified challenges faced by OEC while proposing areas for future research. Edge computing has garnered significant attention in recent years, and the integration of a satellite network with a ground network for edge computing has emerged as a research and development focus for both academia and industry. The feasibility and effectiveness of satellite edge computing have been proven, leading to the design of optimization algorithms for computation offloading and resource management in OEC, as well as the development of testbeds specifically designed for satellite edge computing. OEC has demonstrated wide-ranging potential in diverse fields, including space technology, agricultural remote sensing, the military, the Internet of Things (IoT), and next-generation mobile communication. In the future, the development of OEC will enable significant changes in edge computing and facilitate the growth of the IoT.

## Declaration of Competing Interest

The authors declare that they have no known competing financial interests or personal relationships that could have appeared to influence the work reported in this paper.

**Acknowledgements**

The project was funded by the Hong Kong-Macau-Taiwan Science and Technology Cooperation Project of the Science and Technology Innovation Action Plan in Shanghai, China (23510760200), the Oriental Talent Youth Program of Shanghai, China (No.Y3DFRCZL01), the Outstanding Program of the Youth Innovation Promotion Association of the Chinese Academy of Sciences (No.Y2023080), and the Strategic Priority Research Program of the Chinese Academy of Sciences - Category A (No.XDA0360404).

**References**

1. Tang QQ, Fei ZS, Li B, et al. Computation offloading in LEO satellite networks with hybrid cloud and edge computing. *IEEE Internet Things J* 2021;**8**(11):9164–76.
2. Liu JJ, Shi YP, Fadlullah ZM, et al. Space-air-ground integrated network: a survey. *IEEE Commun Surv Tutor* 2018;**20**(4):2714–41.
3. Henri Y. The OneWeb satellite system. *Handbook of small satellites: technology, design, manufacture, applications, economics and regulation*. Berlin: Springer; 2020. p. 1091–100.
4. Foreman VL, Siddiqi A, De Weck O. Large satellite constellation orbital debris impacts: case studies of oneweb and spacex proposals. *AIAA SPACE and astronautics forum and exposition*. 2017, AIAA; Reston.
5. Project kuiper [Internet]; 2023. Available from: https://www.aboutamazon.com/what-we-do/devices-services/project-kuiper.
6. Telesat: global satellite operators [Internet]; 2020. Available from: https://www.telesat.com/.
7. Patidar S, Rane D, Jain P. A survey paper on cloud computing. *2012 Second international conference on advanced computing & communication technologies*. Piscataway: IEEE Press; 2012. p. 394–8.
8. Mao YY, You CS, Zhang J, et al. A survey on mobile edge computing: the communication perspective. *IEEE Commun Surv Tutor* 2017;**19**(4):2322–58.
9. Wang SG, Li Q, Xu MW, et al. Tiansuan constellation: an open research platform. *2021 IEEE international conference on edge computing (EDGE)*. Piscataway: IEEE Press; 2021. p. 94–101.
10. George AD, Wilson CM. Onboard processing with hybrid and reconfigurable computing on small satellites. *Proc IEEE* 2018;**106**(3):458–70.
11. Wang SG, Li Q. Satellite computing: vision and challenges. *IEEE Internet Things J* 2023;**10**(24):22514–29.
12. Kim T, Choi JP. Performance analysis of satellite server mobile edge computing architecture. *2020 IEEE 92nd vehicular technology conference (VTC2020-Fall)*. Piscataway: IEEE Press; 2020. p. 1–6.
13. Zhang L, Zhang HL, Guo C, et al. Satellite-aerial integrated computing in disasters: user association and offloading decision. *ICC 2020–2020 IEEE international conference on communications (ICC)*. Piscataway: IEEE Press; 2020. p. 554–9.
14. Jiang YH, Li XY, Zhu GX, et al. 6G non-terrestrial networks enabled low-altitude economy: opportunities and challenges. arXiv preprint: 2311.09047; 2023.
15. Oakes B, Richards D, Barr J, et al. Double deep Q networks for sensor management in space situational awareness. *2022 25th international conference on information fusion (FUSION)*. Piscataway: IEEE Press; 2022. p. 1–6.
16. Li CC, Zhang YS, Xie RC, et al. Integrating edge computing into low earth orbit satellite networks: architecture and prototype. *IEEE Access* 2021;**9**:39126–37.
17. Denby B, Lucia B. Orbital edge computing: machine inference in space. *IEEE Comput Archit Lett* 2019;**18**(1):59–62.
18. Denby B, Chintalapudi K, Chandra R, et al. Kodan: addressing the computational bottleneck in space. *28th ACM international conference on architectural support for programming languages and operating systems (ASPLOS '2023)*. New York: ACM; 2023. p. 392–403.
19. Pfandzelter T, Bermbach D. Celestial: virtual software system testbeds for the LEO edge. *23rd ACM/IFIP international middleware conference (Middleware '22)*. New York: ACM; 2022. p. 69–81.
20. Wei JY, Han JR, Cao SZ. Satellite IoT edge intelligent computing: a research on architecture. *Electronics* 2019;**8**(11):1247.
21. Wang F, Jiang DD, Qi S, et al. Fine-grained resource management for edge computing satellite networks. *2019 IEEE global communications conference (GLOBECOM)*. Piscataway: IEEE Press; 2019. p. 1–6.
22. Fang J, Liu HT, Sun JY, et al. Learning-based task offloading in dynamic orbital edge computing network. *2021 IEEE 23rd Int Conf on high performance computing & communications*. Piscataway: IEEE Press; 2021. p. 495–502.
23. Liu SJ, Hu X, Wang WD. Deep reinforcement learning based dynamic channel allocation algorithm in multibeam satellite systems. *IEEE Access* 2018;**6**:15733–42.
24. Altaf N. The next frontier of innovation [Internet]; 2019. Available from: https://www.nasa.gov/wp-content/uploads/2019/10/space_portal_naeem_altaf.pdf.
25. Edge network computing capabilities for satellite remote terminals [Internet]; 2024. Available from: https://connectivity.esa.int/projects/edgesat.
26. Shen ZS, Jin J, Tan C, et al. A survey of next-generation computing technologies in space-air-ground integrated networks. *ACM Comput Surv* 2023;**56**(1):1–40.
27. Li XT, Xu S, Zhao ZP, et al. A survey on computing offloading in satellite-terrestrial integrated edge computing networks. *2023 15th international conference on communication software and networks (ICCSN)*. Piscataway: IEEE Press; 2023. p. 172–82.
28. Centenaro M, Costa CE, Granelli F, et al. A survey on technologies, standards and open challenges in satellite IoT. *IEEE Commun Surv Tutor* 2021;**23**(3):1693–720.
29. Denby B, Lucia B. Orbital edge computing: nanosatellite constellations as a new class of computer system. *Proceedings of the 25-th international conference on architectural support for programming languages and operating systems (ASPLOS)*. New York: ACM; 2020. p. 939–54.
30. Shi WS, Cao J, Zhang Q, et al. Edge computing: vision and challenges. *IEEE Internet Things J* 2016;**3**(5):637–46.
31. Gost MM, Leyva-Mayorga I, Pérez-Neira A, et al. Edge computing and communication for energy-efficient Earth surveillance with LEO satellites. *2022 IEEE international conference on communications workshops (ICC Workshops)*. Piscataway: IEEE Press; 2022. p. 556–61.
32. Leyva-Mayorga I, Martinez-Gost M, Moretti M, et al. Satellite edge computing for real-time and very-high resolution Earth observation. *IEEE Trans Commun* 2023;**71**(10):6180–94.
33. Lucia B, Denby B, Manchester Z, et al. Computational nanosatellite constellations: opportunities and challenges. *GetMobile: Mobile Comp and Comm* 2021;**25**(1):16–23.
34. Guo Y, Wang SG. Challenges and opportunities in space service computing. *2021 IEEE international conference on services computing (SCC)*. Piscataway: IEEE Press; 2021. p. 44–51.
35. Luo MS, Feng JL. *Satellite communications*. Beijing: Tsinghua University Publishing House; 2020. p. 12–21 [Chinese].
36. Al-Hraishawi H, Chougrani H, Kisseleff S, et al. A Survey on nongeostationary satellite systems: the communication perspective. *IEEE Commun Surv Tutor* 2023;**25**(1):101–32.

37. Maral G, Bousquet M, Sun ZL. Satellite communications systems: systems, techniques and technology. Hoboken: Wiley; 2020. p. 12–7.
38. Fossa CE, Raines RA, Gunsch GH, et al. An overview of the IRIDIUM (R) low Earth orbit (LEO) satellite system. In: *Proc IEEE 1998 national aerospace and electronics conference. NAECON 1998*. Piscataway: IEEE Press; 1998. p. 152–9.
39. Elbert BR. *Introduction to satellite communication*. New York: Artech house; 2008. p. 419–25.
40. Su YT, Liu YQ, Zhou YQ, et al. Broadband LEO satellite communications: architectures and key technologies. *IEEE Wirel Commun* 2019;**26**(2):55–61.
41. Wall M. SpaceX Says Its 60 Starlink satellites are all phoning home (and fading out) [Internet]; 2019. Available from: https://www.space.com/spacex-starlink-satellites-phone-home-dimming.html.
42. Lu CY, Cao Y, Peng CZ, et al. Micius quantum experiments in space. *Rev Mod Phys* 2022;**94**(3):035001.
43. Al-Hourani A, Guvenc I. On modeling satellite-to-ground path-loss in urban environments. *IEEE Commun Lett* 2021;**25**(3):696–700.
44. Matthiesen B, Razmi N, Leyva-Mayorga I, et al. Federated learning in satellite constellations. *IEEE Netw* 2024;**38**(2):232–9.
45. Cao SZ, Wei JY, Han H, et al. Space edge cloud enabling network slicing for 5G satellite network. *2019 15th international wireless communications & mobile computing conference (IWCMC)*. Piscataway: IEEE Press; 2019. p. 787–92.
46. Chen ZK, Xiong W. Analysis of satellite communication network characteristics. *2014 4-th International conference on communication systems and network technologies*. Piscataway: IEEE Press; 2014. p. 317–20.
47. Nguyen DC, Ding M, Pathirana PN, et al. 6G Internet of things: a comprehensive survey. *IEEE Internet Things J* 2021;**9**(1):359–83.
48. Ahmed E, Rehmani MH. Mobile edge computing: opportunities, solutions, and challenges. *Futur Gener Comp Syst* 2017;**70**:59–63.
49. Patel M, Naughton B, Chan C, et al. Mobile-edge computing - introductory technical white paper [Internet]; 2014. Available from: https://portal.etsi.org/portals/0/tbpages/mec/docs/mobile-edge_computing_-_introductory_technical_white_paper_v1%2018-09-14.pdf.
50. Pallas F, Raschke P, Bermbach D. Fog computing as privacy enabler. *IEEE Internet Comput* 2020;**24**(4):15–21.
51. Shi WS, Dustdar S. The promise of edge computing. *Computer* 2016;**49**(5):78–81.
52. Yates D, Islam MZ. Data mining on smartphones: an introduction and survey. *ACM Comput Surv* 2022;**55**(5):1–38.
53. Barry B, Brick C, Connor F, et al. Always-on vision processing unit for mobile applications. *IEEE Micro* 2015;**35**(2):56–66.
54. Rapuano E, Meoni G, Pacini T, et al. An FPGA-based hardware accelerator for CNNs inference on board satellites: benchmarking with myriad 2-based solution for the CloudScout case study. *Remote Sens* 2021;**13**(8):1518.
55. Handley M. Delay is not an option: low latency routing in space. *Proceedings of the 17th ACM workshop on hot topics in networks (HotNets)*. New York: ACM; 2018. p. 85–91.
56. Kassar M, Kervella B, Pujolle G. An overview of vertical handover decision strategies in heterogeneous wireless networks. *Comput Commun* 2008;**31**(10):2607–20.
57. Zhang Z, Zhang W, Tseng FH. Satellite mobile edge computing: improving QoS of high-speed satellite-terrestrial networks using edge computing techniques. *IEEE Netw* 2019;**33**(1):70–6.
58. Wang C, Zhang YR, Li Q, et al. Satellite computing: a case study of cloud-native satellites. *2023 IEEE international conference on edge computing and communications (EDGE)*. Piscataway: IEEE Press; 2023. p. 262–70.
59. Toyoshima M. Recent trends in space laser communications for small satellites and constellations. *J Lightwave Technol* 2021;**39**(3):693–9.
60. Chaudhry AU, Yanikomeroglu H. Laser intersatellite links in a starlink constellation: a classification and analysis. *IEEE Veh Technol Mag* 2021;**16**(2):48–56.
61. Saeed N, Elzanaty A, Almorad H, et al. CubeSat communications: recent advances and future challenges. *IEEE Commun Surv Tutor* 2020;**22**(3):1839–62.
62. Guelman M, Kogan A, Kazarian A, et al. Acquisition and pointing control for inter-satellite laser communications. *IEEE Trans Aerosp Electron Syst* 2004;**40**(4):1239–48.
63. Huang QQ, Lin M, Zhu WP, et al. Uplink massive access in mixed RF/FSO satellite-aerial-terrestrial networks. *IEEE Trans Commun* 2021;**69**(4):2413–26.
64. Ansari IS, Yilmaz F, Alouini MS. Impact of pointing errors on the performance of mixed RF/FSO dual-hop transmission systems. *IEEE Wirel Commun Letters* 2013;**2**(3):351–4.
65. Arienzo L. Green RF/FSO communications in cognitive relay-based space information networks for maritime surveillance. *IEEE Trans Cogn Commun Netw* 2019;**5**(4):1182–93.
66. Zhang B, Wu YF, Zhao B, et al. Progress and challenges in intelligent remote sensing satellite systems. *IEEE J Sel Top Appl Earth Observ Remote Sens* 2022;**15**:1814–22.
67. Wu YZ, Xiao LX, Zhou JX, et al. Large-scale MIMO enabled satellite communications: concepts, technologies, and challenges. *IEEE Commun Mag* 2024;**62**(8):140–6.
68. You L, Li KX, Wang JH, et al. Massive MIMO transmission for LEO satellite communications. *IEEE J Sel Areas Commun* 2020;**38**(8):1851–65.
69. Abdelsadek MY, Kurt GK, Yanikomeroglu H. Distributed massive MIMO for LEO satellite networks. *IEEE Open J Commun Soc* 2022;**3**:2162–77.
70. Xu JL, Zhang CJ, Wang CH, et al. Approach to inter-satellite time synchronization for micro-satellite cluster. *J Syst Eng Electron* 2018;**29**(4):805–15.
71. Zhang YY, Wu YL, Liu AJ, et al. Deep learning-based channel prediction for LEO satellite massive MIMO communication system. *IEEE Wirel Commun Letters* 2021;**10**(8):1835–9.
72. Yang XN, Xu JL, Lou CY. Software-defined satellite: a new concept for space information system. *2012 2-th international conference on instrumentation, measurement, computer, communication and control*. Piscataway: IEEE Press; 2012. p. 586–9.
73. Xie RC, Tang QQ, Wang QN, et al. Satellite-terrestrial integrated edge computing networks: architecture, challenges, and open issues. *IEEE Netw* 2020;**34**(3):224–31.
74. Li KT, Hofmann CA, Reder H, et al. A techno-economic assessment and tradespace exploration of low earth orbit mega-constellations. *IEEE Commun Mag* 2023;**61**(2):24–30.
75. Yu S, Gong XW, Shi Q, et al. EC-SAGINs: edge-computing-enhanced space-air-ground-integrated networks for Internet of vehicles. *IEEE Internet Things J* 2022;**9**(8):5742–54.
76. Cheng N, Quan W, Shi WS, et al. Slicing-based task offloading in space-air-ground integrated vehicular networks. *IEEE Trans Mob Comput* 2024;**23**(5):4009–24.
77. Cheng N, Quan W, Shi WS, et al. A Comprehensive simulation platform for space-air-ground integrated network. *IEEE Wirel Commun* 2020;**27**(1):178–85.
78. Kim T, Kwak J, Choi JP. Satellite edge computing architecture and network slice scheduling for IoT Support. *IEEE Internet Things J* 2022;**9**(16):14938–51.

79. Giuseppe A, Giacomo G, Antonio I, et al. Virtualizing nanosatellites in SDN/NFV enabled ground segments to enhance service orchestration. *2019 IEEE global communications conference (GLOBECOM)*. Piscataway: IEEE Press; 2019. p. 1–6.
80. Grossman E, Gouzman I. Space environment effects on polymers in low Earth orbit. *Nucl Instrum Meth B* 2003;**208**:48–57.
81. Pfandzelter T, Bermbach D. Edge computing in low-Earth orbit – What could possibly go wrong? *Proceedings of the 1st ACM workshop on LEO networking and communication (LEO-NET)*. New York: ACM; 2023. p. 19–24.
82. Saleh R, Wilton S, Mirabbasi S, et al. System-on-chip: reuse and integration. *Proc IEEE* 2006;**94**(6):1050–69.
83. Sheldon D, Gagne J, Daniel A, et al. Radiation effects characterization and system architecture options for the 7nm snapdragon SA8155P automotive grade system on chip (SoC). *2022 22nd European conference on radiation and its effects on components and systems (RADECS)*. Piscataway: IEEE Press; 2022. p. 1–4.
84. Giuffrida G, Fanucci L, Meoni G, et al. The Phi-Sat-1 mission: the first on-board deep neural network demonstrator for satellite Earth observation. *IEEE Trans Geosci Remote Sensing* 2021;**60**:1–14.
85. Bhattacherjee D, Kassing S, Licciardello M, et al. In-orbit computing: an outlandish thought experiment? *Proceedings of the 19th ACM workshop on hot topics in networks (HotNets)*. New York: ACM; 2020. p. 197–204.
86. Xu MW, Zhang L, Li HY, et al. A Satellite-born server design with massive tiny chips towards in-space computing. *2022 IEEE international conference on satellite computing (Satellite)*. New York: ACM; 2022. p. 1–6.
87. Eickhoff J. *Onboard computers, onboard software and satellite operations: an introduction*. Berlin: Springer; 2011. p. 89–93.
88. Leppinen H. Current use of Linux in spacecraft flight software. *IEEE Aerosp Electron Syst Mag* 2017;**32**(10):4–13.
89. Wang YX, Yang J, Guo XY, et al. Satellite edge computing for the Internet of things in aerospace. *Sensors* 2019;**19**(20):4375.
90. Fayyaz M, Tanya V, Caujolle JM. Adaptive middleware design for satellite fault-tolerant distributed computing. *2012 NASA/ESA conference on adaptive hardware and systems (AHS)*. Piscataway: IEEE Press; 2022. p. 23–30.
91. Kothari V, Liberis E, Lane ND. The final frontier: deep learning in space. *Proceedings of the 21st international workshop on mobile computing systems and applications (HotMobile)*. New York: ACM; 2020. p. 45–9.
92. Jia M, Zhang LA, Wu J, et al. Collaborative satellite-terrestrial edge computing network for everyone-centric customized services. *IEEE Netw* 2022;**37**(5):197–205.
93. Hu YF, Gong WB. An on-orbit task-offloading strategy based on satellite edge computing. *Sensors* 2023;**23**(9):4271.
94. Zhang HY, Liu RK, Kaushik A, et al. Satellite edge computing with collaborative computation offloading: an intelligent deep deterministic policy gradient approach. *IEEE Internet Things J* 2023;**10**(10):9092–107.
95. Zhang PY, Chen N, Shen SG, et al. AI-enabled space-air-ground integrated networks: management and optimization. *IEEE Netw* 2024;**38**(2):186–92.
96. Li Q, Wang SG, Ma X, et al. Service coverage for satellite edge computing. *IEEE Internet Things J* 2022;**9**(1):695–705.
97. Zhang QL, Luo Y, Jiang H, et al. Aerial edge computing: a survey. *IEEE Internet Things J* 2023;**10**(16):14357–74.
98. Sabol C, Burns R, McLaughlin CA. Satellite formation flying design and evolution. *J Spacecr Rockets* 2001;**38**(2):270–8.
99. Zhang SB, Liu AJ, Han C, et al. Multi-agent reinforcement learning-based orbital edge offloading in SAGIN supporting Internet of remote things. *IEEE Internet Things J* 2023;**10**(23):20472–83.
100. Bui TA, Lee PJ, Liang CS, et al. Edge-computing-enabled deep learning approach for low-light satellite image enhancement. *IEEE J Sel Top Appl Earth Observ Remote Sens* 2024;**17**:4071–83.
101. Wang Q, Chen XM, Qi Q. Energy-efficient design of satellite-terrestrial computing in 6G wireless networks. *IEEE Trans Commun* 2024;**72**(3):1759–72.
102. Zheng GH, Ni Q, Navaie K, et al. Semantic communication in satellite-borne edge cloud network for computation offloading. *IEEE J Sel Areas Commun* 2024;**42**(5):1145–58.
103. Zhan T, Yin K, Xiong JH, et al. Augmented reality and virtual reality displays: perspectives and challenges. *Iscience* 2020;**23**(8):101397.
104. Jiang XT, Yu FR, Song T, et al. A survey on multi-access edge computing applied to video streaming: some research issues and challenges. *IEEE Commun Surv Tutor* 2021;**23**(2):871–903.
105. Hu M, Luo XZ, Chen JW, et al. Virtual reality: a survey of enabling technologies and its applications in IoT. *J Netw Comput Appl* 2021;**178**:102970.
106. Cheng N, Lyu F, Quan W, et al. Space/aerial-assisted computing offloading for IoT applications: a learning-based approach. *IEEE J Sel Areas Commun* 2019;**37**(5):1117–29.
107. Lin K, Li C, Pace P, et al. Multi-level cluster-based satellite-terrestrial integrated communication in Internet of vehicles. *Comput Commun* 2020;**149**:44–50.
108. Nativi S, Mazzetti P, Santoro M, et al. Big data challenges in building the global Earth observation system of systems. *Environ Modell Softw* 2015;**68**:1–26.
109. He YM, Chen YW, Lu JM, et al. Scheduling multiple agile earth observation satellites with an edge computing framework and a constructive heuristic algorithm. *J Syst Architect* 2019;**95**:55–66.
110. Li DX, Xie WY, Li YS, et al. FedFusion: manifold-driven federated learning for multi-satellite and multi-modality fusion. *IEEE Trans Geosci Remote Sensing* 2024;**62**:1–13.
111. Salim S, Moustafa N, Hassanian M, et al. Deep-federated-learning-based threat detection model for extreme satellite communications. *IEEE Internet Things J* 2024;**11**(3):3853–67.
112. McMahan B, Moore E, Ramage D, et al. Communication-efficient learning of deep networks from decentralized data. *Proceedings of the 20th international conference on artificial intelligence and statistics*. New York: PMLR; 2017. p. 1273–82.
113. Razmi N, Matthiesen B, Dekorsy A, et al. Scheduling for ground-assisted federated learning in LEO satellite constellations. *2022 30th European Signal Processing Conference (EUSIPCO)*. Piscataway: IEEE Press; 2022. p. 1102–6.
114. Chen H, Xiao M, Pang ZB. Satellite-based computing networks with federated learning. *IEEE Wirel Commun* 2022;**29**(1):78–84.
115. Jing Y, Jiang CX, Ge N, et al. Resource optimization for signal recognition in satellite MEC with federated learning. *2021 13th international conference on wireless communications and signal processing (WCSP)*. Piscataway: IEEE Press; 2021. p. 1–5.
116. Elmahallawy M, Luo T. AsyncFLEO: Asynchronous federated learning for LEO satellite constellations with high-altitude platforms. *2022 IEEE international conference on big data (Big Data)*. Piscataway: IEEE Press; 2022. p. 5478–87.
117. So J, Hsieh K, Arzani B, et al. Fedspace: an efficient federated learning framework at satellites and ground stations. arXiv preprint: 2202.01267; 2022.
118. Razmi N, Matthiesen B, Dekorsy A, et al. On-board federated learning for dense LEO constellations. *ICC 2022-IEEE international conference on communications*. Piscataway: IEEE Press; 2022. p. 4715–20.
119. Elmahallawy M, Luo T. FedHAP: Fast federated learning for LEO constellations using collaborative HAPs. *2022 14th international conference on wireless communications and signal processing (WCSP)*. Piscataway: IEEE Press; 2022. p. 888–93.

120. Razmi N, Matthiesen B, Dekorsy A, et al. Ground-assisted federated learning in LEO satellite constellations. *IEEE Wirel Commun Letters* 2022;**11**(4):717–21.
121. Leyva-Mayorga I, Soret B, Popovski P. Inter-plane inter-satellite connectivity in dense LEO constellations. *IEEE Trans Wirel Commun* 2021;**20**(6):3430–43.
122. Han DJ, Hosseinalipour S, Love DJ, et al. Cooperative federated learning over ground-to-satellite integrated networks: joint local computation and data offloading. *IEEE J Sel Areas Commun* 2024;**42**(5):1080–96.
123. Razmi N, Matthiesen B, Dekorsy A, et al. On-board federated learning for satellite clusters with inter-satellite links. *IEEE Trans Commun* 2024;**72**(6):3408–24.
124. Perez-Portero A, Munoz-Martin JF, Park H, et al. Airborne GNSS-R: a key enabling technology for environmental monitoring. *IEEE J Sel Top Appl Earth Observ Remote Sens* 2021;**14**:6652–61.
125. Tang FX, Wen C, Chen XH, et al. Federated learning for intelligent transmission with space-air-ground integrated network (SAGIN) toward 6G. *IEEE Netw* 2022;**37**(2):198–204.
126. Yang MH, Zhang JJ, Liu SY. DFedSat: communication-efficient and robust decentralized federated learning for LEO satellite constellations. arXiv preprint: 2407.05850; 2024.
127. Elmahallawy M, Luo T, Ramadan K. Communication-efficient federated learning for LEO constellations integrated with HAPs using hybrid NOMA-OFDM. *IEEE J Sel Areas Commun* 2024;**42**(5):1097–114.
128. Yang C, Yuan JL, Wu YZ, et al. Communication-efficient satellite-ground federated learning through progressive weight quantization. *IEEE Trans Mob Comput* 2024;**23**(9):8999–9011.
129. Wu CH, He SY, Yin ZS, et al. Towards client selection in satellite federated learning. *Appl Sci-Basel* 2024;**14**(3):1286.
130. Sonmez C, Ozgovde A, Ersoy C. EdgeCloudSim: an environment for performance evaluation of edge computing systems. *Trans Emerg Telecommun Technol* 2018;**29**(11):1102–6.
131. Gupta H, Vahid Dastjerdi A, Ghosh SK, et al. iFogSim: a toolkit for modeling and simulation of resource management techniques in the Internet of things, edge and fog computing environments. *Software: Practice and Experience* 2017;**47** (9):1275–96.
132. Zeng XZ, Garg SK, Strazdins P, et al. IOTSim: a simulator for analysing IoT applications. *J Syst Architect* 2017;**72**:93–107.
133. Wang B, Li XH, Huang DY, et al. A profit maximization strategy of MEC resource provider in the satellite-terrestrial double edge computing system. *2021 IEEE 21st international conference on communication technology (ICCT)*. Piscataway: IEEE Press; 2021. p. 906–12.
134. Song ZY, Hao YY, Liu YW, et al. Energy-efficient multiaccess edge computing for terrestrial-satellite Internet of things. *IEEE Internet Things J* 2021;**8**(18):14202–18.
135. Wang YX, Yang J, Guo XY, et al. A Game-theoretic approach to computation offloading in satellite edge computing. *IEEE Access* 2020;**8**:12510–20.
136. Pfandzelter T, Bermbach D. QoS-aware resource placement for LEO satellite edge computing. *2022 IEEE 6th international conference on fog and edge computing (ICFEC)*. Piscataway: IEEE Press; 2022. p. 66–72.
137. Pfandzelter T, Bermbach D. Edge (of the earth) replication: optimizing content delivery in large LEO satellite communication networks. *2021 IEEE/ACM 21st international symposium on cluster, cloud and internet computing (CCGrid)*. Piscataway: IEEE Press; 2021. p. 565–75.
138. Kim T, Choi JP. Performance analysis of satellite server mobile edge computing architecture. *2020 IEEE 92nd vehicular technology conference (VTC2020-Fall)*. Piscataway: IEEE Press; 2020. p. 1–6.
139. Cui GF, Li XY, Xu LX, et al. Latency and energy optimization for MEC enhanced SAT-IoT networks. *IEEE Access* 2020;**8**:55915–26.
140. Li HF, Chen C, Li C, et al. Aerial computing offloading by distributed deep learning in collaborative satellite-terrestrial networks. *2021 13th international conference on wireless communications and signal processing (WCSP)*. Piscataway: IEEE Press; 2021. p. 1–6.
141. Zhu DL, Liu HT, Li T, et al. Deep reinforcement learning-based task offloading in satellite-terrestrial edge computing networks. *2021 IEEE wireless communications and networking conference (WCNC)*. Piscataway: IEEE Press; 2021. p. 1–7.
142. Christie G, Fendley N, Wilson J, et al. Functional Map of the World. *Proc. IEEE conference on computer vision and pattern recognition (CVPR)*. Piscataway: IEEE Press; 2018. p. 6172–80.
143. Yan ZG, Li D, Yu XH, et al. Latency-efficient wireless federated learning with quantization and scheduling. *IEEE Commun Lett* 2022;**26**(11):2621–5.
144. Puttonen J, Rantanen S, Laakso F, et al. Satellite model for network simulator 3. Workshop on simulation for european space programmes (SESP). Paris: ESA; 2015. p. 86–91.
145. Kassing S, Bhattacherjee D, Águas AB, et al. Exploring the 'Internet from space' with Hypatia. *Proceedings of the ACM internet measurement conference (IMC)*. New York: ACM; 2020. p. 214–29.
146. Kempton B, Riedl A. Network simulator for large low Earth orbit satellite networks. *ICC 2021-IEEE international conference on communications*. Piscataway: IEEE Press; 2021. p. 1–6.
147. Lai ZQ, Li HW, StarPerf Li JH. Characterizing network performance for emerging mega-constellations. *2020 IEEE 28th international conference on network protocols (ICNP)*. Piscataway: IEEE Press; 2020. p. 6172–80.
148. Zhang Y, Wang B, Guo BL, et al. A Research on integrated space-ground information network simulation platform based on SDN. *Comput Netw* 2021;**188**:107821.
149. Abbas N, Zhang Y, Taherkordi A, et al. Mobile edge computing: a survey. *IEEE Internet Things J* 2018;**5**(1):450–65.
150. Wang YJ, Zhang JX, Zhang X, et al. A Computation offloading strategy in satellite terrestrial networks with double edge computing. *2018 IEEE international conference on communication systems (ICCS)*. Piscataway: IEEE Press; 2018. p. 450–5.
151. Ding CF, Wang JB, Zhang H, et al. joint optimization of transmission and computation resources for satellite and high altitude platform assisted edge computing. *IEEE Trans Wirel Commun* 2022;**21**(2):1362–77.
152. Ding CF, Wang JB, Zhang H, et al. Joint MU-MIMO precoding and resource allocation for mobile-edge computing. *IEEE Trans Wirel Commun* 2021;**20**(3):1639–54.
153. Tong ML, Li S, Wang XX, et al. Inter-satellite cooperative offloading decision and resource allocation in mobile edge computing-enabled satellite-terrestrial networks. *Sensors* 2023;**23**(2):668.
154. Wang RS, Zhu WC, Liu GL, et al. Collaborative computation offloading and resource allocation in satellite edge computing. *GLOBECOM 2022–2022 IEEE global communications conference*. Piscataway: IEEE Press; 2022. p. 5625–30.
155. Gao YF, Yan ZB, Zhao KL, et al. Joint optimization of server and service selection in satellite-terrestrial integrated edge computing networks. *IEEE Trans Veh Technol* 2024;**73** (2):2740–54.
156. Hao YY, Song ZY, Zheng Z, et al. Joint communication, computing, and caching resource allocation in LEO satellite MEC networks. *IEEE Access* 2023;**11**:6708–16.

157. Wang HP, Han JR, Cao SZ, et al. Computation offloading strategy of multi-satellite cooperative tasks based on genetic algorithm in satellite edge computing. *2021 international conference on space-air-ground computing (SAGC)*. Piscataway: IEEE Press; 2021. p. 22–8.
158. Panwar P, Shabaz M, Nazir S, et al. Generic edge computing system for optimization and computation offloading of unmanned aerial vehicle. *Comput Electr Eng* 2023;**109**:108779.
159. Wu J, Jia M, Zhang L, et al. DNNs based computation offloading for LEO satellite edge computing. *Electronics* 2022;**11**(24):4108.
160. Shuai JQ, Cui HX, He YJ, et al. Dynamic satellite edge computing offloading algorithm based on distributed deep learning. *IEEE Internet Things J* 2024;**11**(16):27790–802.
161. Chai FR, Zhang Q, Yao HP, et al. Joint multi-task offloading and resource allocation for mobile edge computing systems in satellite IoT. *IEEE Trans Veh Technol* 2023;**72**(6):7783–95.
162. Wang B, Xie JC, Huang DY, et al. A computation offloading strategy for LEO satellite mobile edge computing system. *2022 14th international conference on communication software and networks (ICCSN)*. Piscataway: IEEE Press; 2022. p. 75–80.
163. Lin X, Liu AJ, Han C, et al. LEO satellite and UAVs assisted mobile edge computing for tactical Ad-Hoc network: a game theory approach. *IEEE Internet Things J* 2023;**10**(23):20560–73.
164. Chen ZE, Zhang DY, Cai WJ, et al. A joint resource allocation and task offloading algorithm in satellite edge computing. *International conference on algorithms and architectures for parallel processing (ICA3PP)*. Berlin: Springer; 2023. p. 358–77.
165. Lu CH, Sheu JP, Hsieh CY. An energy optimization algorithm for UAV-assisted satellite mobile edge computing system. *2024 IEEE wireless communications and networking conference (WCNC)*. Piscataway: IEEE Press; 2024. p. 1–6.
166. Fei HX, Zhang X, Long J, et al. Towards multi-satellite collaborative computing via task scheduling based on genetic algorithm. *Aerospace* 2023;**10**(2):95.
167. Mnih V, Kavukcuoglu K, Silver D, et al. Human-level control through deep reinforcement learning. *Nature* 2015;**518**(7540):529–33.
168. Pfandzelter T, Hasenburg J, Bermbach D. Towards a computing platform for the LEO edge. *Proceedings of the 4th international workshop on edge systems, analytics and networking (EdgeSys)*. New York: ACM; 2021. p. 43–8.
169. Jones H. The recent large reduction in space launch cost. *48th international conference on environmental systems (ICES)*. Washington D.C.: NASA; 2021. p. 1–10.
170. Chi CX, Zhang YR, Sun QB, et al. SABM: adaptive backup mechanism for satellite services. *Proceedings of the 1st ACM MobiCom workshop on satellite networking and computing*. New York: ACM; 2023. p. 1–6.
171. Sheng M, Zhou D, Bai WG, et al. Coverage enhancement for 6G satellite-terrestrial integrated networks: performance metrics, constellation configuration and resource allocation. *Sci China Inform Sci* 2023;**66**(3):130303.
172. Valente F, Eramo V, Lavacca FG. Optimal bandwidth and computing resource allocation in low earth orbit satellite constellation for earth observation applications. *Comput Netw* 2023;**232**:109849.
173. Peng C, He YZ, Zhao SH, et al. Integration of data center into the distributed satellite cluster networks: challenges, techniques, and trends. *IEEE Netw* 2023;**37**(3):52–8.
174. Andreev S, Galinina O, Pyattaev A, et al. Exploring synergy between communications, caching, and computing in 5G-grade deployments. *IEEE Commun Mag* 2016;**54**(8):60–9.
175. Han DR, Peng HX, Wu HQ, et al. Joint cache placement and content delivery in satellite-terrestrial integrated C-RANs. *ICC 2021-IEEE international conference on communications*. Piscataway: IEEE Press; 2021. p. 1–6.
176. An K, Li YS, Yan XJ, et al. On the performance of cache-enabled hybrid satellite-terrestrial relay networks. *IEEE Wirel Commun Letters* 2019;**8**(5):1506–9.
177. Zhao RL, Luo JT, Ran YY. Coverage-aware cooperative caching and efficient content distribution schemes in LEO satellite networks. *Proceedings of the 1st ACM mobiCom workshop on satellite networking and computing (SatCom)*. New York: ACM; 2023. p. 31–6.
178. Zhang K, Leng SP, He YJ, et al. Cooperative content caching in 5G networks with mobile edge computing. *IEEE Wirel Commun* 2018;**25**(3):80–7.
179. Wang SG, Guo Y, Zhang N, et al. Delay-aware microservice coordination in mobile edge computing: a reinforcement learning approach. *IEEE Trans Mob Comput* 2021;**20**(3):939–51.
180. Yan L, Cao SZ, Gong YS, et al. SatEC: a 5G satellite edge computing framework based on microservice architecture. *Sensors* 2019;**19**(4):831.
181. Thönes J. Microservices. *IEEE Softw* 2015;**32**(1), 116-6.
182. Huang YZ, Zhang X, Xu ZC. SatEdge: platform of edge cloud at satellite and scheduling mechanism for microservice modules. *IEEE Access* 2023;**11**:126283–94.
183. García-Rois J, Lorenzo B, González-Castaño FJ, et al. Slice allocation and pricing framework for virtualized millimeter wave cellular networks. *IEEE Access* 2019;**7**:86349–66.
184. Esmat HH, Lorenzo B, Shi W. Toward resilient network slicing for satellite-terrestrial edge computing IoT. *IEEE Internet Things J* 2023;**10**(16):14621–45.
185. Nguyen DC, Ding M, Pham QV, et al. Federated learning meets blockchain in edge computing: opportunities and challenges. *IEEE Internet Things J* 2021;**8**(16):12806–25.